\documentclass[preprint, 12pt, authoryear]{elsarticle}
\usepackage[utf8]{inputenc}
\usepackage[english]{babel}
\usepackage{textcomp}

\usepackage{bm}
\newcommand{\vect}[1]{\boldsymbol{#1}}
\usepackage[acronym]{glossaries}
\setacronymstyle{long-short}
\newacronym{leo}{LEO}{Low Earth Orbit}
\newacronym{pnp}{PNP}{Probability of no-penetration}
\newacronym{ble}{BLE}{Ballistic Limit Equation}
\newacronym{drama}{DRAMA}{Debris Risk Assessment and Mitigation Analysis}
\newacronym{esa}{ESA}{European Space Agency}
\newacronym{iadc}{IADC}{Inter-Agency Space Debris Coordination Committee}
\newacronym{nasa}{NASA}{National Aeronautics and Space Administration}
\newacronym{srl}{SRL}{Schafer-Ryan-Lambert}
\newacronym{master}{MASTER}{Meteoroid and Space Debris Terrestrial Environment Reference}
\newacronym{midas}{MIDAS}{MASTER-based Impact Flux and Damage Assessment Software}
\newacronym{cfrp}{CFRP}{Carbon Fibre Reinforced Plastics}
\newacronym{hcsp}{HC-SP}{Honeycomb Sandwich Panel}
\newacronym{sso}{SSO}{Sun-synchronous orbit}
\newacronym{pirat}{PIRAT}{Particle Impact Risk and Vulnerability Analysis Tool}
\usepackage[noprefix]{nomencl}
\makenomenclature
\RequirePackage{ifthen}
\renewcommand{\nomgroup}[1]{%
\ifthenelse{\equal{#1}{V}}{\item[\textbf{Variables}]}{%
\ifthenelse{\equal{#1}{L}}{\item[\textbf{Subscripts}]}{%
\ifthenelse{\equal{#1}{H}}{\item[\textbf{Superscripts}]}{%
\ifthenelse{\equal{#1}{C}}{\item[\textbf{Constants}]}}}}{}}
\newcommand{\nomunit}[1]{%
\renewcommand{\nomentryend}{\hspace*{\fill}#1}}
\usepackage{amsthm, mathtools, amsmath, gensymb, wasysym, array}
\usepackage{siunitx}

\usepackage{tabularx}
\usepackage{url}
\usepackage{hyperref}
\usepackage[capitalise]{cleveref}
\crefrangelabelformat{equation}{(#3#1#4-#5#2#6)}
\usepackage[dvipsnames]{xcolor}
\usepackage{graphicx}
\usepackage{placeins, afterpage}
\usepackage{setspace}
\usepackage{multirow}

\journal{Advances in Space Research}

\begin{document}

\begin{frontmatter}
\title{Predicting the vulnerability of spacecraft components: modelling debris impact effects through vulnerable-zones}
\author{Mirko Trisolini\corref{cor1}\fnref{fn1}}
\ead{mirko.trisolini@polimi.it}
\address{University of Southampton, University Road, SO17 1BJ, Southampton, United Kingdom}
\author{Hugh G. Lewis}
\ead{H.G.Lewis@soton.ac.uk}
\address{University of Southampton, University Road, SO17 1BJ, Southampton, United Kingdom}
\author{Camilla Colombo}
\ead{camilla.colombo@polimi.it}
\address{Politecnico di Milano, Via La Masa 34, 20156, Milano, Italy}
\cortext[cor1]{Corresponding author}
\fntext[fn1]{Present address: Politecnico di Milano, Via La Masa 34, 20156, Milano, Italy}

\begin{abstract}
The space environment around the Earth is populated by more than 130 million objects of 1 mm in size and larger, and future predictions shows that this amount is destined to increase, even if mitigation measures are implemented at a far better rate than today. These objects can hit and damage a spacecraft or its components. It is thus necessary to assess the risk level for a satellite during its mission lifetime. Few software packages perform this analysis, and most of them employ time-consuming ray-tracing methodology, where particles are randomly sampled from relevant distributions. In addition, they tend not to consider the risk associated with the secondary debris clouds. The paper presents the development of a vulnerability assessment model, which relies on a fully statistical procedure: the debris fluxes are directly used combining them with the concept of vulnerable zone, avoiding the random sampling the debris fluxes. A novel methodology is presented to predict damage on internal components. It models the interaction between the components and the secondary debris cloud through basic geometrical operations, considering mutual shielding and shadowing between internal components. The methodologies are tested against state-of-the-art software for relevant test cases, comparing results on external structures and internal components.
\end{abstract}


\end{frontmatter}

\nomenclature[V]{$Az$}{Impact azimuth angle}
\nomenclature[V]{$El$}{Impact elevation angle}
\nomenclature[V]{$v_p$}{Impact velocity of a debris particle}
\nomenclature[V]{$\varphi$}{Debris flux}
\nomenclature[V]{$d_p$}{Particle diameter}
\nomenclature[V]{$d_{c, b}$}{Critical diameter in the ballistic regime}
\nomenclature[V]{$d_{c, h}$}{Critical diameter in the hypervelocity regime}
\nomenclature[V]{$\rho_{p}$}{Particle density}
\nomenclature[V]{$t_w$}{Thickness of the rear wall}
\nomenclature[V]{$t_{ob}$}{Thickness of the outer bumper plate}
\nomenclature[V]{$t_{b}$}{Thickness of the bumper plate}
\nomenclature[V]{$\theta$}{Impact angle}
\nomenclature[V]{$K_{3S}$}{\gls{srl} \gls{ble} ballistic fitting factor}
\nomenclature[V]{$\delta$}{\gls{srl} \gls{ble} ballistic fitting factor}
\nomenclature[V]{$\sigma_{y}$}{Yield strength of the rear wall material}
\nomenclature[V]{$K_{tw}$}{\gls{srl} \gls{ble} hypervelocity fitting factor}
\nomenclature[V]{$K_{S2}$}{\gls{srl} \gls{ble} hypervelocity fitting factor}
\nomenclature[V]{$S_1$}{Spacing between the outer bumper and the bumper plates}
\nomenclature[V]{$S_2$}{Spacing between the bumper plate and the rear wall}
\nomenclature[V]{$\rho_{ob}$}{Density of the outer bumper plate}
\nomenclature[V]{$\beta$}{\gls{srl} \gls{ble} hypervelocity fitting factor}
\nomenclature[V]{$\gamma$}{\gls{srl} \gls{ble} hypervelocity fitting factor}
\nomenclature[V]{$\epsilon$}{\gls{srl} \gls{ble} hypervelocity fitting factor}
\nomenclature[V]{$\Delta T$}{Mission lifetime in years}
\nomenclature[V]{$S$}{Cross-sectional area}
\nomenclature[V]{$\varphi_{c}$}{Critical particle flux}
\nomenclature[V]{$P_{imp}$}{Impact probability}
\nomenclature[V]{$P_{pen}$}{Penetration probability}
\nomenclature[V]{$N_{panels}$}{Number of panels forming the considered structure/object}
\nomenclature[V]{$N_{fluxes}$}{Number of vector flux elements schematising the debris environment}
\nomenclature[V]{$V_{comp}$}{Vulnerability of an internal component}
\nomenclature[V]{$P_{struct}$}{Probability of space debris hitting the spacecraft external structure in a point belonging to a vulnerable zone}
\nomenclature[V]{$P_{cloud}$}{Probability that the secondary cloud ejecta will hit an internal component}
\nomenclature[V]{$P_{BLE}$}{Probability that the particles in the cloud ejecta perforates the component wall}
\nomenclature[V]{$\theta_1$}{Deflection angle of the normal debris cloud}
\nomenclature[V]{$\theta_2$}{Deflection angle of the inline debris cloud}
\nomenclature[V]{$\eta_1$}{Half-cone angle of the normal debris cloud}
\nomenclature[V]{$\eta_2$}{Half-cone angle of the inline debris cloud}
\nomenclature[V]{$C$}{Speed of sound}
\nomenclature[V]{$t_s$}{Thickness of a plate}
\nomenclature[V]{$l_{vz}$}{Extent of the vulnerable zone at the target plane}
\nomenclature[C]{$\alpha_{max}$}{Maximum ejection angle \nomunit{$63.15\, \deg$}}
\nomenclature[V]{$d_{target}$}{Lateral extent of the target component \nomunit{$s$}}
\nomenclature[V]{$d_{p,max}$}{Maximum size of the particle diameter}
\nomenclature[V]{$S_{vz}$}{Projected area of the vulnerable zone}
\nomenclature[V]{$d_{ejecta}$}{Lateral extent of the debris ejecta at the target plane}
\nomenclature[V]{$s$}{Stand-off distance between a component and the external wall}
\nomenclature[V]{$\alpha$}{Secondary fragments ejection angle}
\nomenclature[V]{$\eta$}{Half-aperture angle of the secondary cloud ejecta cone}
\nomenclature[V]{$A_{c,av}$}{Available cone area: difference between the intersection of the cone area at the target plane with the perspective projections of the shielding components}
\nomenclature[V]{$A_{c}$}{Intersection between the debris cone with the target plane}
\nomenclature[V]{$A_{s}$}{Perspective projection of a shielding component onto the target plane}
\nomenclature[V]{$N_{cell}$}{Number of cells subdividing the vulnerable zone}
\nomenclature[V]{$N_{s}$}{Number of shielding components between the target component and the vulnerable zone}
\nomenclature[V]{$\bar{A}_{c,av}$}{Available cone area averaged over the grid cells subdividing a vulnerable zone}
\nomenclature[V]{$A_{t}$}{Area of the target component at the target plane}
\nomenclature[V]{$\bar{A}_{t,av}$}{Target visible area averaged over the grid cells subdividing a vulnerable zone}
\nomenclature[V]{$A_{p}$}{Cross-section of a debris particle. A spherical shape is assumed for the particle}
\nomenclature[V]{$\bar{d}_{t,av}$}{Equivalent extension of the target visible area averaged over the grid cells subdividing a vulnerable zone}
\nomenclature[V]{$\bar{d}_{c,av}$}{Equivalent extension of the available cone area averaged over the grid cells subdividing a vulnerable zone}
\nomenclature[V]{$\bar{A}_{vz,av}$}{Available vulnerable zone area averaged over the grid cells subdividing a vulnerable zone}
\nomenclature[V]{$\bar{d}_{vz,av}$}{Available vulnerable zone extent averaged over the grid cells subdividing a vulnerable zone}
\nomenclature[V]{$\rho_m$}{Material density}
\nomenclature[V]{$HB$}{Material Brinell hardness}

\printnomenclature[1.5cm]

\section{Introduction}
\label{sec:intro}
Since the launch of Sputnik 1 in 1957, the space around the Earth and beyond has been the theatre of remarkable achievements but has also suffered from continuous exploitation. Decommissioned satellites, spent upper stages, mission-related objects, and fragments generated by collisions and explosions of satellites and upper stages have polluted the space environment in the form of space debris. Space debris is thus considered as a major threat to space mission; in fact, debris of just 1 cm can cause the break-up of a satellite \citep{Putzar2006, Stokes2005}, and smaller particles can still have enough energy to produce failures on components critical to the mission operations and scientific objectives. Recent studies have shown a constant increase in the population of space debris, and the amount of debris is expected to keep growing in the next years \citep{Radtke2017, Lewis2017, Lewis2017a, Lewis2012}. The major space-faring nations and international committees have thus proposed a series of debris mitigation measures to protect the space environment \citep{OConnor2008, Schafer2005, IADC2007}. The implementation of these mitigation measures can help to mitigate the impact of human activities on the space environment.

Nonetheless, the danger posed by possible impacts of space debris on spacecraft structures and components has to be addressed and regarded as a mission design driver as a single impact can compromise an entire mission \citep{Grassi2014, Stokes2012, Christiansen2009}. This is particularly important for missions in specific orbital regions, such as highly inclined orbits, which are more densely populated by space debris and impacts can be more frequent \citep{Liou2006}. The space debris population accounts for about 29000 objects larger than 10 cm, 750000 objects between 1 cm to 10 cm, and 166 million objects from 1 mm to 1 cm \citep{SpaceDebrisOffice2017}. In \gls{leo} orbits, only particles larger than 10 cm can be tracked \citep{Schaub2015}. Consequently, only for these particles collision avoidance manoeuvres are effective in preventing damage to satellites. For the other untraceable yet dangerous debris, it is necessary to adopt specific measures to ensure the mission survivability. To do so we need to assess the amount of damage caused by space debris to spacecraft components and identify the most vulnerable ones.
Few software packages are currently available to perform the vulnerability assessment of a mission. Examples are ESABASE2/DEBRIS \citep{Gade2013}, \gls{nasa} BUMPER \citep{Bjorkman2014}, \gls{esa} \gls{drama} \citep{Gelhaus2014,Gelhaus2013, Martin2005, Martin2005a}, SHIELD \citep{Stokes2000, Stokes2005, Stokes2012}, and \gls{pirat} \citep{Gulde2016, Kempf2013, Kempf2016}. These codes have a common structure, where the satellite is modelled through a geometric representation (which can be more or less detailed). The geometry is usually schematised through a set of panels. Each panel is then assigned the properties and characteristics required for the impact analysis, such as the material, the shielding configuration, the thickness, etc. \citep{Grassi2014, Gade2013, Stokes2000, Welty2013}. An environmental model (e.g. \gls{master}) is then used to generate the debris fluxes affecting the spacecraft. The nature and frequency of the impacts on the structure and components of the spacecraft is evaluated and the damage associated with them is computed using semi-empirical relationships (i.e. Ballistic Limit Equations). Finally, the results of the analysis can be obtained in the form of number of impacts, number of penetrations, impact probability, and penetration probability.

These software packages perform the vulnerability analysis with different levels of detail and techniques. The two main aspects to be considered are the modelling of the satellite geometry and the computation of the damage on internal components. For instance, \gls{drama} is only used for preliminary analysis. It only allows the definition of simple geometrical shapes through a limited number of panels, which are created specifying their area, normal, and shielding properties. Also, the impact analysis only involves the outer structure of the satellite and does not consider internal components. BUMPER instead generates a high-fidelity Finite Element Model (FEM) of the spacecraft structure; however, it cannot take into account the impacts on internal components. SHIELD and ESABASE2/DEBRIS both use a panelised model of the satellite structure, and a ray-tracing methodology to simulate the impact of debris particle on the spacecraft. They both can compute the effects of debris impacts on internal components. The methodologies used by the two models for this task are, however, very different. In SHIELD, after the first impact on the external structure, the projectile is considered still intact and following the same direction of the initial impact. In ESABASE2/DEBRIS the assessment for internal components is not directly supported and requires a workaround: the user has to create a panelized representation of the internal components, without the external structure. The contribution of the external structure is then taken into account assigning a multiple-wall \gls{ble} to each panel constituting the internal components. A recent integration to this methodology has been proposed and is under development \citep{Bunte2017}, which considers the characteristics of the debris cloud that is formed inside the spacecraft after the first impact. In this methodology, the characteristic distributions relative to the debris cloud velocity, mass, and spatial distribution are obtained from available experimental data. The distributions are then sampled to generate rays representing the cloud fragments. This type of methodology is, however, time consuming as it requires the generation of thousands of rays and the verification of the impact with the components for each one of them. Besides, as the rays are generated through sampling, several simulations must be performed to have statistically meaningful results. Finally, the tool \gls{pirat} by the Fraunhofer Institute, also relies on a complete 3D modelling of the satellite structure and internal components and on the use of environmental models such as \gls{master} and BUMPER. However, differently from SHIELD and ESABASE2/DEBRIS, uses a deterministic approach instead of a probabilistic one \citep{Kempf2016,Gulde2016,Welty2013}. This approach is based on a three-wall geometrical assessment to define \emph{threat directions}, which generate the \gls{ble} configuration for the components. In particular, it uses the \gls{srl}-\gls{ble}, so that shielding and partial shielding can be taken into account, as well as primary and secondary impacts. It has also been applied in the Concurrent Design Facility at \gls{esa} for preliminary vulnerability analyses.
\\ \\
The work presented in this paper is devoted to the introduction of a novel methodology to assess the penetration probability on internal spacecraft components in an intrinsically statistical fashion \citep{Welty2013} by exploiting and extending the concept of vulnerable zones \citep{Putzar2006}. The need for ray tracing methodology is removed and so is the need for several Monte Carlo runs to obtain robust results. Also, the implementation of vulnerable zones is coupled with the formation of secondary ejecta clouds after the first impact. The methodology relies then on geometrical operations to compute the interactions of the debris cloud with the internal components. The presented methodology thus evolves from SHIELD, where the particle remains intact, by considering the cloud contribution, and avoids computationally expensive ray-tracing procedure as in \citet{Bunte2017}. The methodology has been developed as a part of a broader study \citep{Trisolini2018_Acta, Trisolini2018_AESCTE, Trisolini2016_JSSE, Trisolini2015_Jerusalem}, where a fast assessment of the vulnerability of simplified spacecraft configurations was required. We first describe the overall model developed, for then focusing on the specifics of the computation of the vulnerability of internal components. The results are then compared with \gls{drama} and ESABASE2/DEBRIS for relevant test cases.

\section{Model outline}
\label{sec:outline}
To properly describe the developed methodology, it is important to understand the entire vulnerability model in all its parts. The model evaluates the vulnerability of simplified spacecraft configurations, considering the effects on both the external structure and on the internal components \citep{Trisolini2018_Acta, Trisolini2018_AESCTE}. It takes as inputs the spacecraft configuration, the mission scenario, and the debris fluxes from a space environment model, and returns the impact and penetration probability on the external structure and internal components. The main novelties in the model are the schematization of the debris environment through \emph{vector flux elements} (Section \ref{subsubsec:vfe}), the implementation of a statistical methodology to compute the impact and penetration probabilities through the use of vulnerable zones (Section \ref{subsec:vulnerable_zone}), and the development of a fully geometrical methodology for the assessment of the mutual shielding between internal components (Section \ref{subsec:mutual_shielding}) without relying on ray tracing methodologies.
\subsection{Spacecraft configuration}
\label{subsec:sc_config}
The spacecraft configuration is defined through a set of elementary shapes such as boxes, cylinders, spheres, and panels. For each element in the configuration, its main properties must be provided, such as the shape, the size, the mass, and the position inside the spacecraft. An example of the configuration entries is provided in Table \ref{tab:config}; the main body of the satellite is always the first entry and is positioned at the centre of the reference frame. The \textit{x}-axis points towards the direction of the orbital velocity (RAM direction), the \textit{z}-axis points away from the Earth, and the \textit{y}-axis follows from the right-hand rule.

\begin{table}[hbt!]
\caption{\label{tab:config} Example of spacecraft configuration structure used by the model.}
\centering
{\renewcommand{\arraystretch}{1.1}%
\begin{tabular}{llcccccccc}
\hline
\textbf{ID} & \textbf{Name} & \textbf{Shape} & \textbf{m} & \textbf{l} & \textbf{r} & \textbf{w} & \textbf{h} & \textbf{Pos.} \\
 &  &  & \textbf{(kg)} & \textbf{(m)} & \textbf{(m)} & \textbf{(m)} & \textbf{(m)} & \textbf{(m)} \\ \hline
0 & Spacecraft & Box & 2000 & 3.5 & n/a & 1.5 & 1.5 & n/a \\
1 & Tank & Sphere & 15 & n/a & 0.55 & n/a & n/a & \text{(-1,0,0)} \\
2 & BattBox & Box & 5 & 0.6 & n/a & 0.5 & 0.4 & \text{(0,0,1)} \\
 \text{...} & \text{...} & \text{...} & \text{...} & \text{...} & \text{...} & \text{...} & \text{...} & \text{...} \\
\hline
\end{tabular}}
\end{table}
The satellite structure can be defined with a single material and a uniform thickness; alternatively, different materials, thickness, and shielding types can be assigned to each face. Three types of shielding can be used in the current version of the model: single wall shields, Whipple shields, and honeycomb sandwich panels \citep{Ryan2010}. For the internal components, the position must be specified; the components can be free inside the main structure, or they can be attached to the external panels. The orientation of a component can also be specified, but only along the main axes of the satellite (\textit{x}, \textit{y}, and \textit{z}). Once the geometry of the spacecraft has been finalised, the software extracts all the relevant information needed for the vulnerability assessment, such as the shield type, the stand-off distances, the vulnerable zones extents, etc. Using this information together with the vector flux elements (Section \ref{subsubsec:vfe}), and the ballistic limit equations (Section \ref{subsec:ballistic}), the vulnerability of each internal component can be computed.
\subsection{Space environment model}
\label{subsec:space_env}
\begin{figure}[!htb]
\centering
\includegraphics[width=0.6\textwidth]{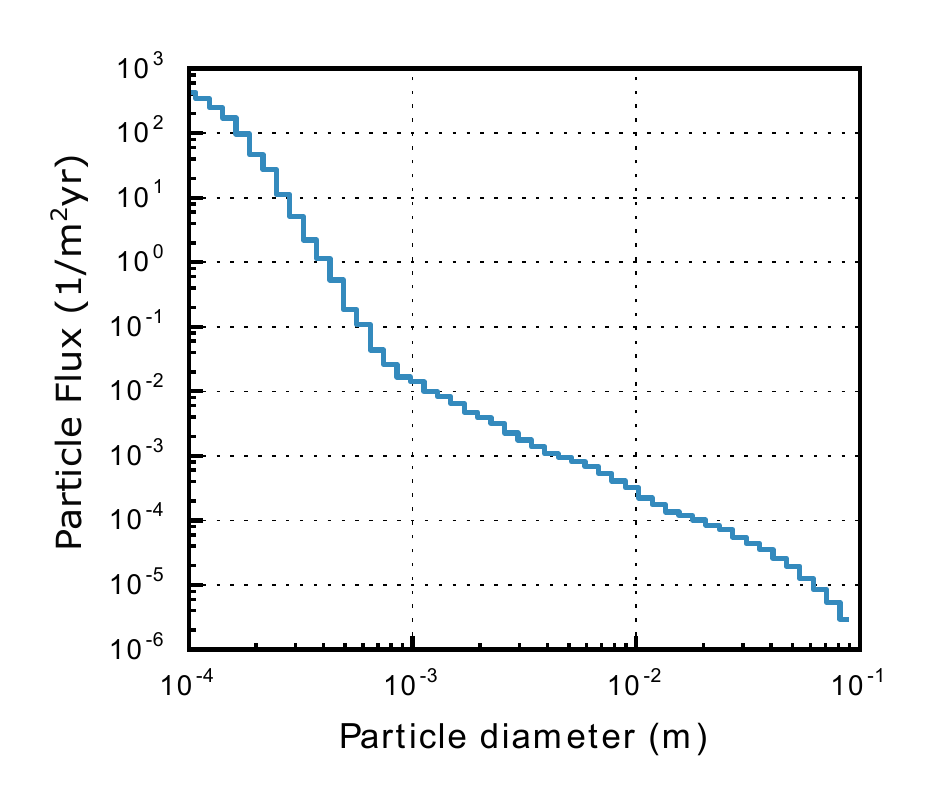}
	\caption{Example of \gls{master}-2009 cumulative flux vs particle diameter distribution. Orbit characteristics: $a$ = 7185 km, $e$ = 0.001, $i$ = 98$^\circ$. Epoch: 1st May 2009. Mission lifetime: 1 year}
	\label{fig:dia-flux}
\end{figure}

\begin{figure}[!htb]
\centering
\includegraphics[width=0.6\textwidth]{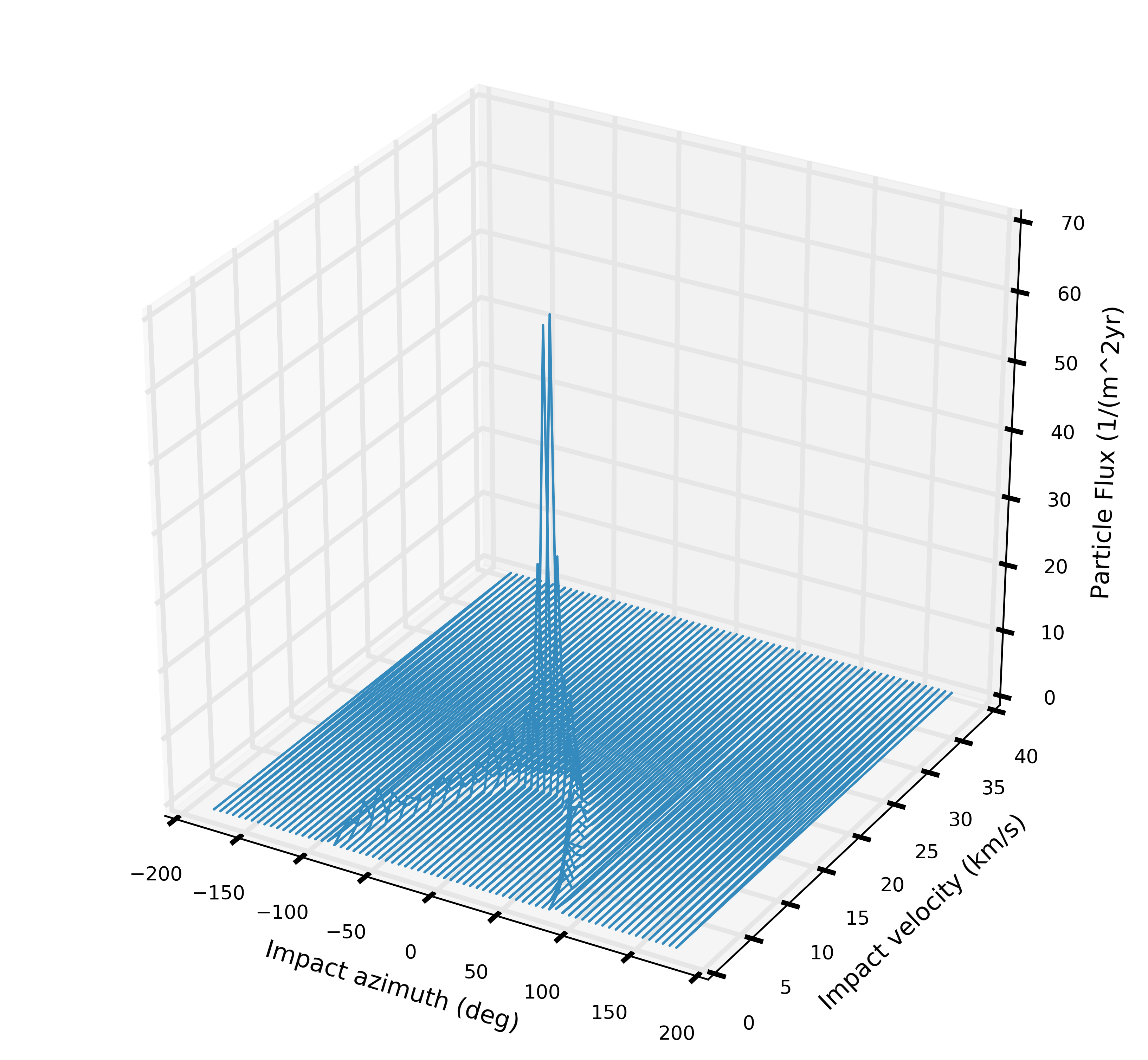}
	\caption{Example of \gls{master}-2009 differential flux vs impact azimuth vs impact velocity distribution.}
	\label{fig:vel_azi}
\end{figure}

To assess the impact of space debris on a satellite, it is necessary to know the characteristics of the environment the satellite has to face during its lifetime. The characteristics and number of debris particles impacting a spacecraft are dependent upon the mission scenario, mainly its altitude and inclination. The present model uses \gls{esa} \gls{master} 2009 \citep{Flegel_MASTER_Report, Gelhaus2011master} to obtain the description of the debris environment via flux predictions on user-defined target orbits \citep{Trisolini2018_Acta}. The fluxes provide the number of particles per unit area per year that impact the spacecraft, given the specific mission scenario (i.e. orbital parameters). The distributions needed for the survivability computation are the flux vs particle diameter, flux vs impact elevation, flux vs impact elevation vs impact azimuth, and the flux vs impact velocity vs impact azimuth. This set of distributions is sufficient to perform the spacecraft vulnerability analysis. \cref{fig:vel_azi} and \ref{fig:dia-flux} represent an example of the flux distributions given by \gls{master}, where the differential flux as a function of the impact velocity and impact azimuth and the cumulative flux as a function of the particle diameter are presented. The fluxes refer to a \gls{sso} with semi-major axis of 7185 km, eccentricity of 0.001, and inclination of 98$^\circ$ for one year with starting epoch 1st May 2009.

\subsubsection{Vector Flux Elements}
\label{subsubsec:vfe}
\begin{figure}[htb!]
\centering
\includegraphics[height=8cm, keepaspectratio]{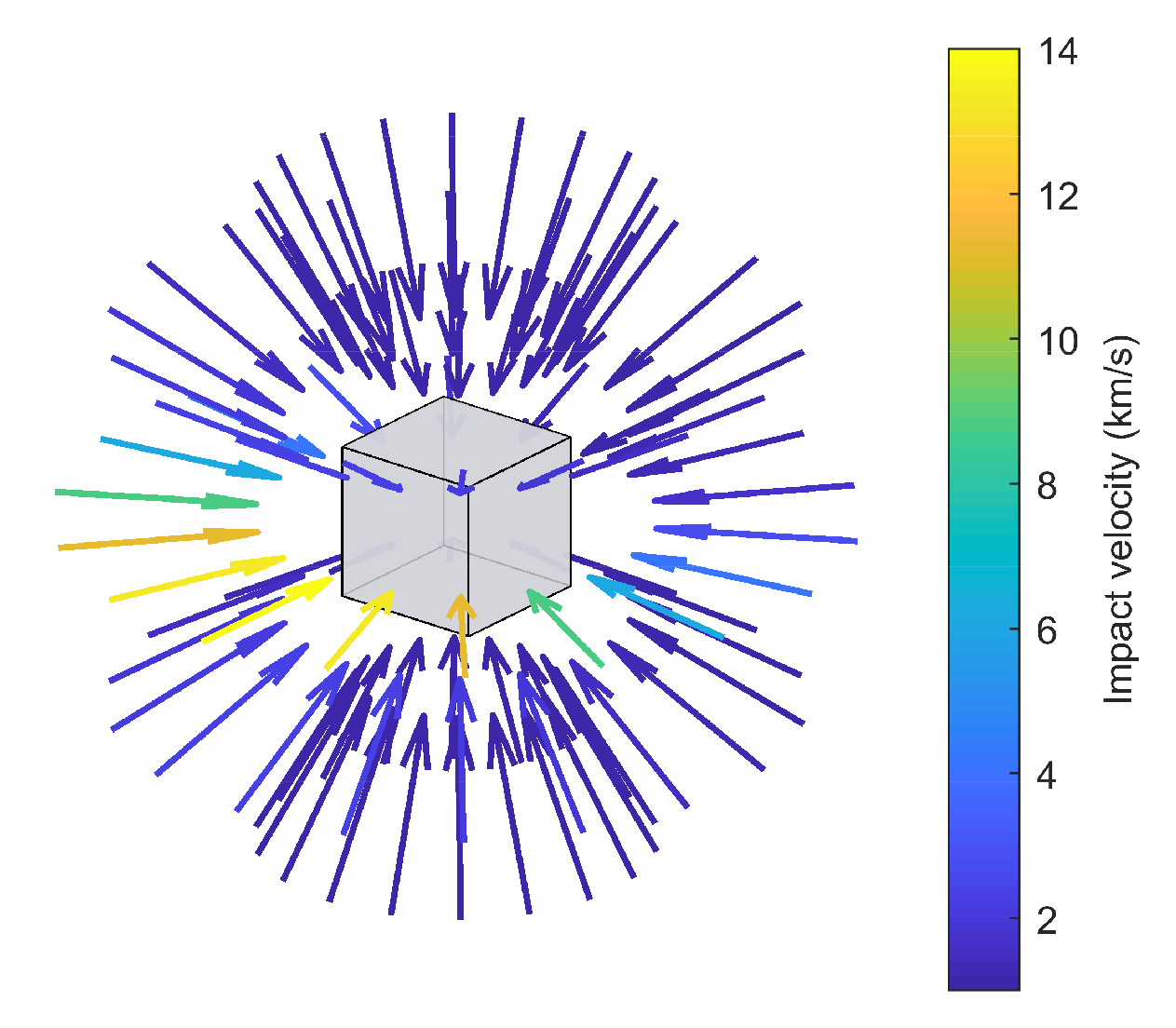}
	\caption{Vector flux elements representation. In this example each vector carries the value of the impact velocity for the angular sector it is referred to.}
	\label{fig:vfe}
\end{figure}

Despite \gls{master}-2009 provides the description of the debris environment through fluxes, this information cannot be directly used in the computation of the vulnerability of a spacecraft. In fact, assessing the criticality of the impact of a debris fragment requires the knowledge of the particle diameter, velocity, and direction. These data can be obtained by directly sampling the flux distributions provided by \gls{master} to generate a set of \emph{impacting particles}, which are shot towards the satellite. This procedure is at the core of ray tracing methods, which are computationally expensive and require a large number of simulations in order to be statistically robust \citep{Stokes2000, Stokes2005}.
The methodology implemented in this work, instead, follows a novel approach to bypass the use of such computationally expensive procedures, and at the same time have a sufficiently accurate description of the debris environment. The compromise is achieved by characterising the debris environment surrounding the spacecraft with \emph{vector flux elements}.
The role of vector flux elements is to summarise the information relative to a specific angular sector of space around the satellite (expressed in terms of impact azimuth and elevation) as provided by \gls{master}-2009. In fact,  each vector element is associated with a value of the particle flux, particle diameter, impact velocity, and impact direction (\cref{fig:vfe}) and these values are then used for the entire sector.

The complete procedure to generate the vector flux elements is described in detail in \citet{Trisolini2018_Acta}. In summary, it consists of subdividing the space around the satellite into a set of angular sectors of impact azimuth and elevation. Then, for each angular sector, the diameter, the impact velocity, and the impact direction are extracted from the \gls{master}-2009 distributions. As in each angular sector these quantities have a distribution and not a single value, we need to use a criterion to select the value to be associated to the \emph{vector flux element}. The criterion we use is the weighted average of the quantity of interest, using as weights the corresponding values of the particle fluxes. For a generic quantity \textit{x}, inside the angular sector $[Az_i,\,Az_{i+1}; El_i,\,El_{i+1}]$ we thus have

\begin{equation} \label{eq:weighted_avg}
\bar{x}_{ij} = \frac{\sum_{k=1}^{M} x_k\varphi_k}{\sum_{k=1}^{M} \varphi_k} \qquad i \in \{ 1,...,N_{Az} \}; \:\: j \in \{ 1,...,N_{El} \};
\end{equation}

where $\bar{x}_{ij}$ is the weighted average value of the quantity \textit{x} referred to the \textit{ij} angular sector, $\varphi_k$ is the value of the flux corresponding to the quantity $x_k$, which is obtained from a histogram equivalent to the one of \cref{fig:dia-flux} but limited to the considered angular sector, and \textit{M} is the number of subdivisions used to obtain the flux histogram from \gls{master}-2009. $N_{Az}$, and $N_{El}$ are the number of angular sector subdivision for the azimuth and the elevation angle respectively. Using this procedure for all the quantities of interest we obtain a vector in the form of \cref{eq:vfe_formula}, which represents all the properties defining a vector flux element referred to an angular sector.

\begin{equation} \label{eq:vfe_formula}
\vect{f}_{ij} = \begin{bmatrix} \bar{\varphi}_{ij} \;\, \bar{v}_{p,ij} \;\, \bar{Az}_{ij} \;\, \bar{El}_{ij} \;\, \bar{d}_{p, ij} \end{bmatrix}
\end{equation}

The use of the schematisation introduced by the \emph{vector flux elements} aims at simplifying the computation and have a computationally efficient procedure. In fact, with this, we can limit the number of times we check for penetration using BLEs and, in particular, the number of geometrical operations that are required to assess the contribution of the clouds associated to the secondary ejecta. A better precision could be achieved directly using the binned output generated by \gls{master}-2009. However, this would come at the cost of a higher computational time.

\subsection{Ballistic Limit Equations}
\label{subsec:ble}
When debris particles hit a spacecraft, the type of impact and its effects need to be assessed. Such an assessment is performed using a set of experimentally derived analytical expressions referred to as Ballistic Limit Equations (\glspl{ble}) \citep{Ryan2010, Ryan2011, Christiansen2009, Schafer2005}. Given the impact speed, the impact angle, the projectile density, and the target characteristics, \glspl{ble} allow the computation of the critical particle diameter, which is the minimum diameter an impacting particle must have to produce damage on the specified target. In general, \glspl{ble} have different expressions as a function of the type of impact. Three different types of impacts are distinguished depending on the relative impact velocity: \emph{ballistic}, \emph{hypervelocity}, and \emph{shatter} regime \citep{Ryan2010, Ryan2008, Ryan2011}. The expressions for the \glspl{ble} are further divided by the shielding type: single-wall, double-wall, triple-wall, and advanced shielding concepts all have dedicated expressions \citep{Ryan2010}. Single-wall \glspl{ble} are subdivided according to the material of the shield, from classical metallic materials such as aluminium, titanium, and stainless steel, to \gls{cfrp}, fibreglass, glass, and polycarbonate. Dual-wall shields include two main configurations: the standard metallic Whipple shield \citep{Ryan2010}, and the \gls{hcsp} \citep{Ryan2008}. A commonly used triple-wall \gls{ble} for the analysis of spacecraft vulnerability is the \gls{srl} equation \crefrange{eq:srl_ball}{eq:srl_hyp} \citep{Schafer2008, Ryan2008, Putzar2006, Grassi2014, Welty2013}. The \gls{srl} \gls{ble} allows the computation of the critical diameter for equipment placed inside the main structure of a satellite by considering the last plate of the triple-wall configuration as the face of the target equipment \citep{Putzar2006}. As this \gls{ble} is versatile and can be used for internal components, we will use it for the reminder of the work. The expression of the \gls{srl} \gls{ble} for the ballistic regime ($v_p \leq V_{LV}\cdot\cos{\theta}$) is

\begin{equation} \label{eq:srl_ball}
d_{c,b} = \Bigg[ \frac{\frac{1}{K_{3S}} \cdot \big( t_w^{0.5}+t_b \big) \cdot \Big( \frac{\sigma_y}{40} \Big)^{0.5} + t_{ob}}{0.6 \cdot \big( \cos{\theta} \big)^{\delta} \cdot \rho_{p}^{0.5}\cdot v_{p}^{2/3}} \Bigg]^{18/19}
\end{equation}

where $v_p$ is the particle velocity relative to the spacecraft, $\rho_p$ is the particle density, $\theta$ is the impact angle (angle between the relative velocity vector and the normal to the impacted surface), $V_{LV}$ is the ballistic regime transition velocity, $K_{3S}$ and $\delta$ are fitting factors whose values are summarised in Table \ref{tab:srl_coeff}, $t_b$ is the bumper plate thickness, $t_{ob}$ is the outer bumper thickness, $t_w$ is the rear wall thickness, and $\sigma_{y}$ is the yield strength of the rear wall material. In the hypervelocity regime ($v_p \geq V_{HV}\cdot\cos{\theta}$) instead we have

\begin{equation} \label{eq:srl_hyp}
d_{c,h} = \frac{1.155 \cdot \bigg[ S_{1}^{1/3} \cdot \big( t_b + K_{tw} \cdot t_w \big)^{2/3} + K_{S2}\cdot S_{2}^{\beta} \cdot t_{w}^{\gamma} \big( \cos{\theta} \big)^{-\epsilon} \bigg] \cdot \bigg( \frac{\sigma_y}{70} \bigg)^{1/3}}{K_{3D}^{2/3} \cdot \rho_{p}^{1/3} \cdot \rho_{ob}^{1/9} \cdot v_{p}^{2/3} \cdot \big( \cos{\theta} \big)^{\delta}}
\end{equation}

where $V_{HV}$ is the hypersonic regime transition velocity. $K_{tw}$, $K_{S2}$, $K_{3D}$, $\beta$, $\epsilon$, and $\gamma$ are fitting factors (Table \ref{tab:srl_coeff}); $\rho_{ob}$ is the outer bumper density; $S_1$ and $S_2$ are the spacing between the outer bumper and the bumper plate, and the space between the bumper plate and the rear wall respectively. Finally, in the shatter regime ($V_{LV}\cdot\cos{\theta} \leq v_p \leq V_{HV}\cdot\cos{\theta}$) a linear interpolation is used between the critical diameters obtained in the ballistic and in the hypervelocity regimes.
The \gls{srl} \gls{ble} has been used throughout this work for the computation of the critical diameter for impacts on components insides the spacecraft. Single-wall and dual-wall \glspl{ble} can also be used by the model to predict impact damage on external structures and components.

\subsection{Impact and penetration probability assessment}
\label{subsec:prob_assessment}
\begin{figure}[htb!]
\centering
\includegraphics[width=0.55\textwidth, keepaspectratio]{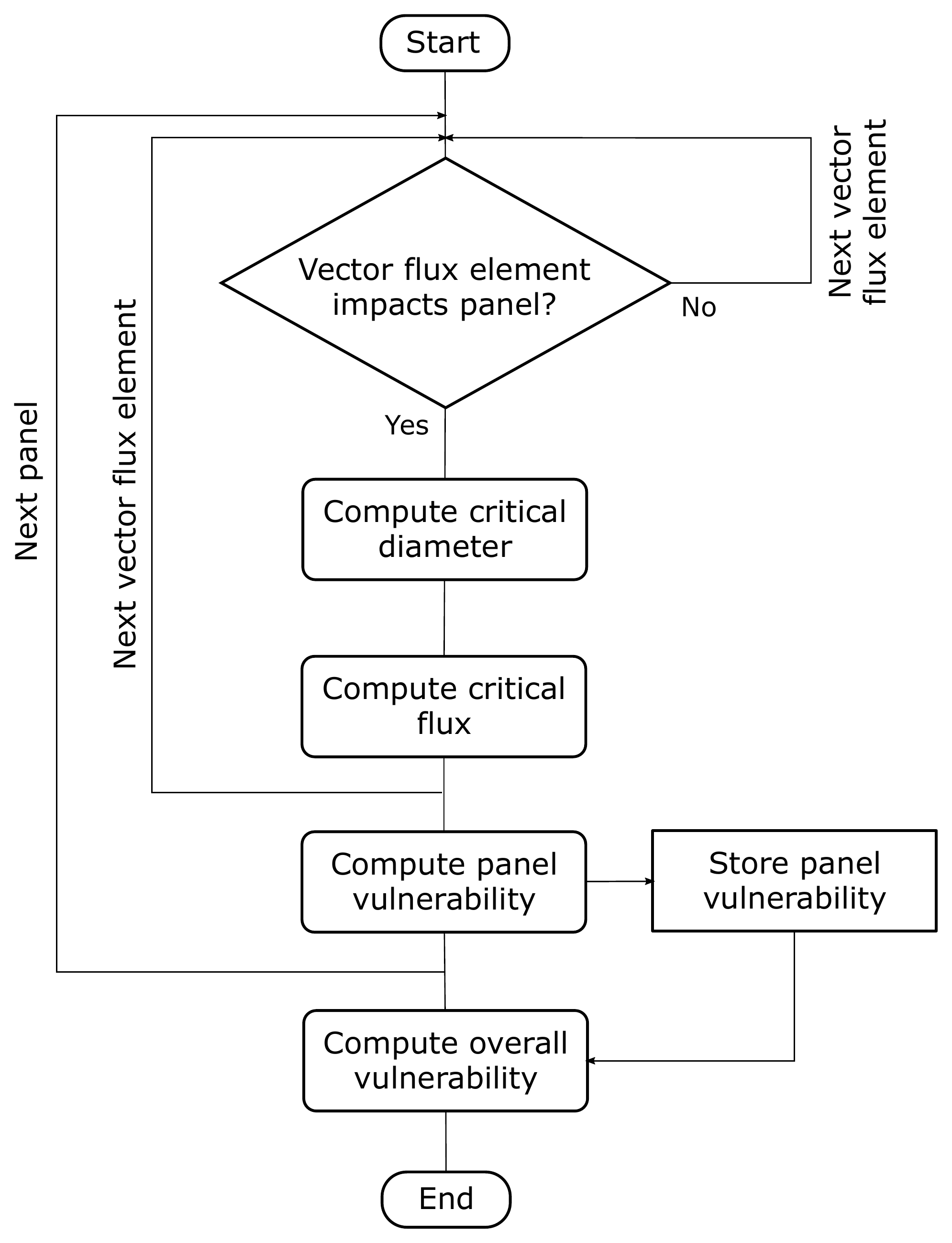}
	\caption{Flow diagram for the vulnerability computation.}
	\label{fig:diagram_vulnerability}
\end{figure}

The procedure for the computation of the impact probability ($P_{imp}$) and penetration probability ($P_{pen}$) depends on the characteristics of the environment surrounding the satellite that are the particle flux, velocity, direction, and diameter. In addition, the orientation of the surface considered plays an important role as it influences the area that is \emph{visible} by a specific \emph{vector flux element}. Finally, even the mission lifetime is an important parameter to be considered as the more a spacecraft resides in a specific debris environment, the more it will be exposed to the debris fluxes. The outline of the procedure is presented in \cref{fig:diagram_vulnerability} and the detailed description can be found in \citet{Trisolini2018_AESCTE}.

To evaluate the impact and penetration probabilities, the structure of the spacecraft is schematised as a set of panels. For each panel, it is necessary to check which vector flux element can actually impact the panel. This is achieved using a simple visibility which checks that $\vect{v_i} \cdot \vect{n_j} < 0$, where $\vect{v_i}$ is the \textit{i}-th vector flux element and $\vect{n_j}$ is the normal to the \textit{j}-th panel in the set. If a vector flux element impacts the panel, its contribution to the overall impact probability is then evaluated. By using Poisson statistics that is by assuming that debris impact events are statistically independent, the impact probability of a vector flux element on an a panel can be expressed as

\begin{equation} \label{eq:imp_prob}
P_{imp}^{ij} = 1 - \exp{\Big( -\varphi_i \cdot S_{ij} \cdot \Delta T} \Big)
\end{equation}

where $\varphi_i$ is the flux associated to the \textit{i}-th vector flux element, $S_{ij}$ is the projected area of the \textit{j}-th face on the direction of the \textit{i}-th vector flux element, and $\Delta T$ is the mission time in years.
For the computation of the penetration probability, the classic approach involving \glspl{ble} can be adopted \citep{Kuiper2010, Reimerdes2001, Welty2013, Stokes2000, Grassi2014, Bunte2009, Stokes2012}. Substituting into the relevant \gls{ble} the velocity and direction of the vector flux elements, the critical diameters are computed. The critical diameter is then used to compute the corresponding \emph{critical flux} i.e. the particle flux corresponding to diameters greater than the critical diameter.

\begin{figure}[htb!]
\centering
\includegraphics[width=0.65\textwidth, keepaspectratio]{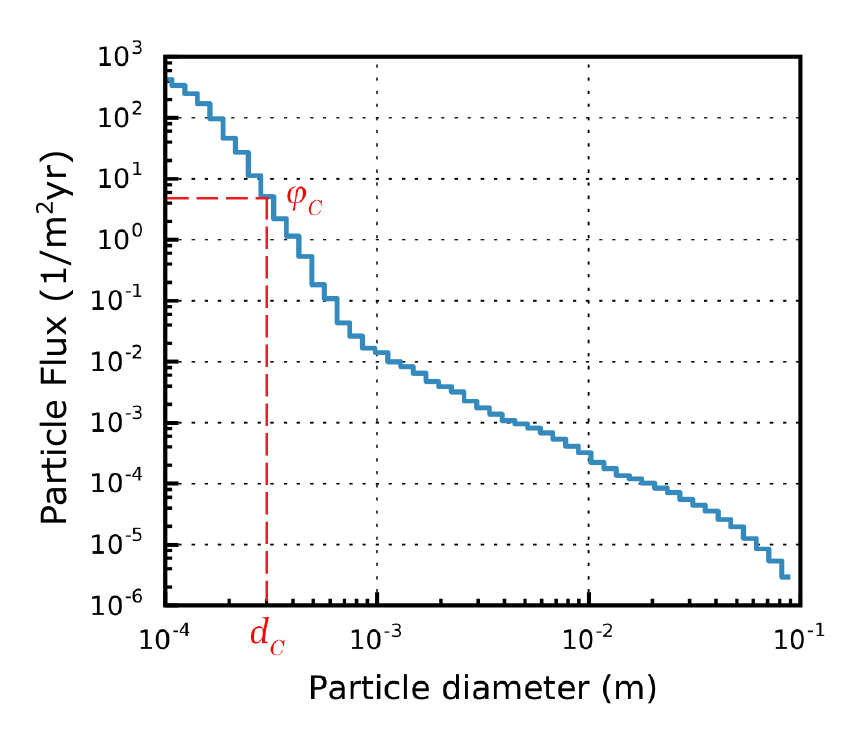}
	\caption{Graphical representation of the methodology for the computation of the critical flux. Orbit characteristics: $a$ = 7185 km, $e$ = 0.001, $i$ = 98$^\circ$. Epoch: 1st May 2009. Mission lifetime: 1 year}
	\label{fig:critical-flux}
\end{figure}

To compute the critical flux, we use the \gls{master}-2009 distribution of the cumulative flux vs particle diameter, from where the it can be extracted (\cref{fig:critical-flux}). Finally, the expression for the penetration probability is given by

\begin{equation} \label{eq:pen_prob}
P_{pen}^{ij} = 1 - \exp{\Big( -\varphi_{c,i} \cdot S_{ij} \cdot \Delta T} \Big)
\end{equation}

where $\varphi_{c,i}$ is the critical flux associated to the \textit{i}-th vector flux element. 

\begin{figure}[htb!]
\centering
\includegraphics[width=0.65\textwidth, keepaspectratio]{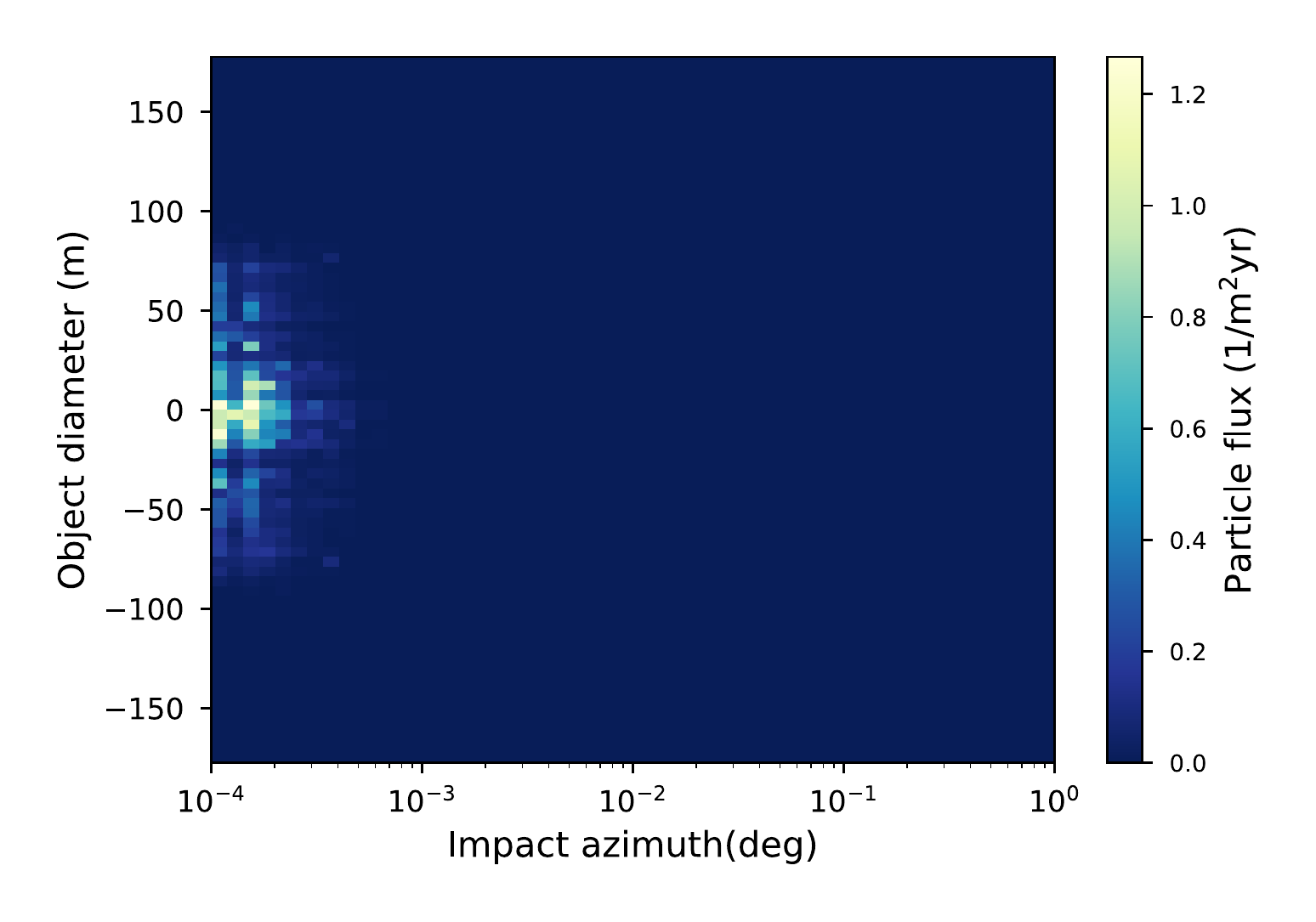}
	\caption{Particle flux vs Impact elevation vs Object diameter. Orbit characteristics: $a$ = 7185 km, $e$ = 0.001, $i$ = 98$^\circ$. Epoch: 1st May 2009. Mission lifetime: 1 year}
	\label{fig:azi_diam}
\end{figure}

\begin{figure}[htb!]
\centering
\includegraphics[width=0.65\textwidth, keepaspectratio]{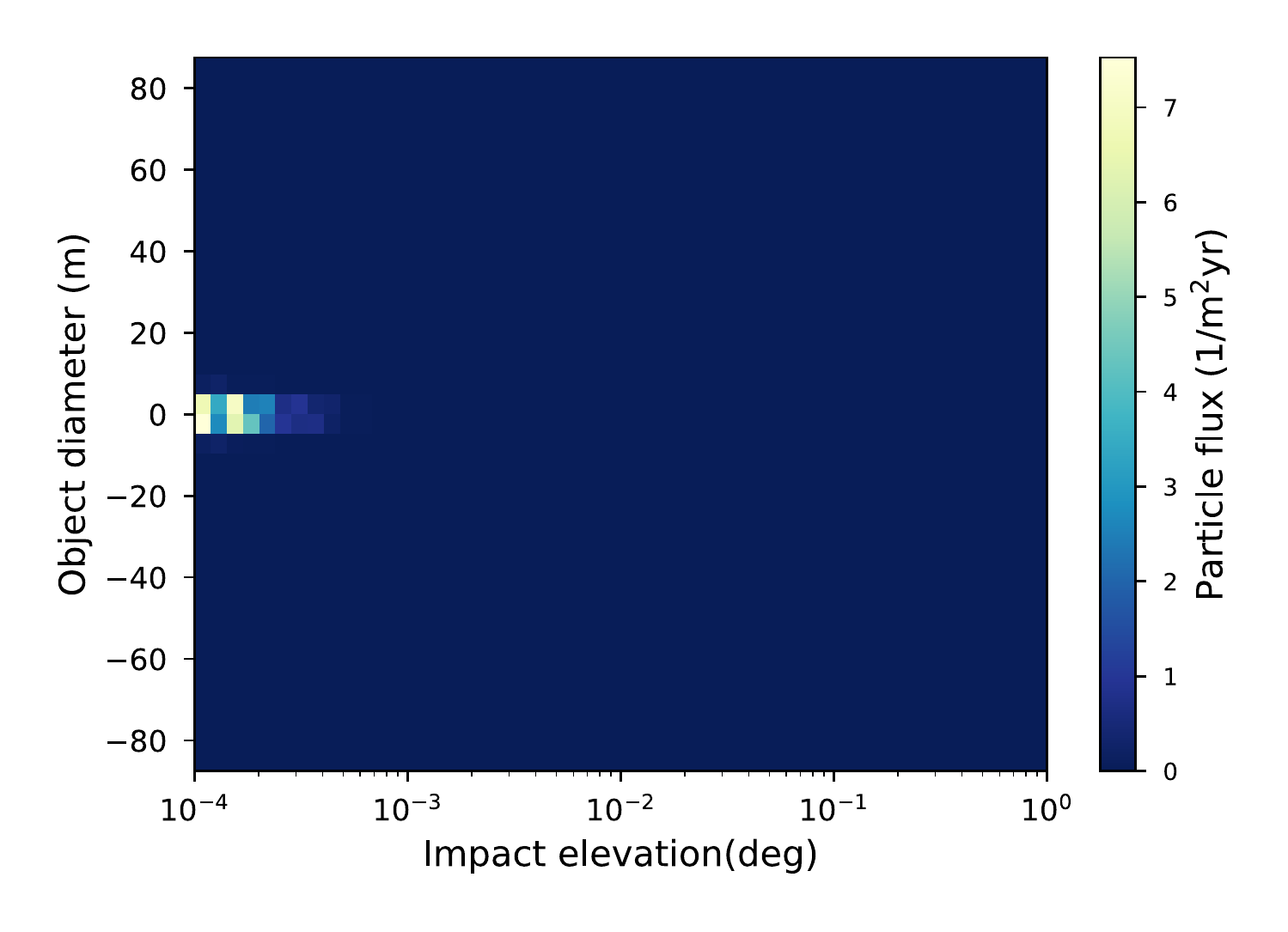}
	\caption{Particle flux vs Impact elevation vs Object diameter. Orbit characteristics: $a$ = 7185 km, $e$ = 0.001, $i$ = 98$^\circ$. Epoch: 1st May 2009. Mission lifetime: 1 year}
	\label{fig:ele_diam}
\end{figure}

It is important to highlight the simplification introduced with this procedure: as the global distribution of cumulative flux vs diameter is used for computing the critical diameter, the obtained flux is the overall flux for the entire range of azimuth and elevation angles. This flux cannot be directly used to compute the penetration probability associated to a single vector flux element as to each one of them is associated a value of the particle flux that is dependent upon the directionality, i.e. impact elevation and impact azimuth. It is thus assumed that the shape of the curve describing the relation between the particle flux and the particle diameter maintains the same shape through the different values of impact azimuth and impact elevation. By doing so, we obtain the critical flux for a vector flux element as a fraction of the total critical flux, which is proportional to the flux associated to that particular vector flux element. With this assumption, the critical flux associated to a vector flux element can be computed as follows

\begin{equation} \label{eq:critical_flux}
\varphi_{c,i} = \varphi_i \cdot \frac{\varphi_c}{\varphi_{tot}}
\end{equation}

where $\varphi_{tot}$ is the total debris flux and $\varphi_c$ is the overall critical flux as extracted from \cref{fig:critical-flux}. \cref{fig:azi_diam,fig:ele_diam} show the distribution of the particle flux for the particle diameter and the impact azimuth and elevation respectively. From these maps can be better observed that the assumption is not always satisfied: there are regions where the inverse exponential trend of the overall particle flux with the diameter (\cref{fig:critical-flux}) is maintained. In these regions, the scaling assumption can be reasonable; however, it is less accurate for other regions. The simplified nature of the proposed method is aligned with the introduced simplification; nonetheless, future development of the methodology will revise this assumption, including a more accurate determination of the critical flux.

Finally, the overall impact and penetration probabilities (\cref{eq:Pimp,eq:Ppen} can be computed iterating over the entire set of vector flux elements and spacecraft panels as follows:

\begin{equation} \label{eq:Pimp}
P_{imp} = 1 - \prod_{j=1}^{N_{panels}} \Bigg( \prod_{i=1}^{N_{fluxes}} \big( 1 - P_{imp}^{ij} \big) \Bigg)
\end{equation}

\begin{equation} \label{eq:Ppen}
P_{pen} = 1 - \prod_{j=1}^{N_{panels}} \Bigg( \prod_{i=1}^{N_{fluxes}} \big( 1 - P_{pen}^{ij} \big) \Bigg)
\end{equation}

where $N_{fluxes}$ is the total number of vector flux elements and $N_{panels}$ is the total number of panels composing the structure considered. The outlined procedure is general and can be used for both the external structure of the spacecraft and for the internal components, provided the relevant changes are taken into account. \cref{sec:vulnerability} provides a detailed description of such corrections with the complete procedure for the assessment of the vulnerability of internal components.

\section{Vulnerability of internal components}
\label{sec:vulnerability}
\begin{figure}[htb!]
\centering
\includegraphics[width=0.6\textwidth, keepaspectratio]{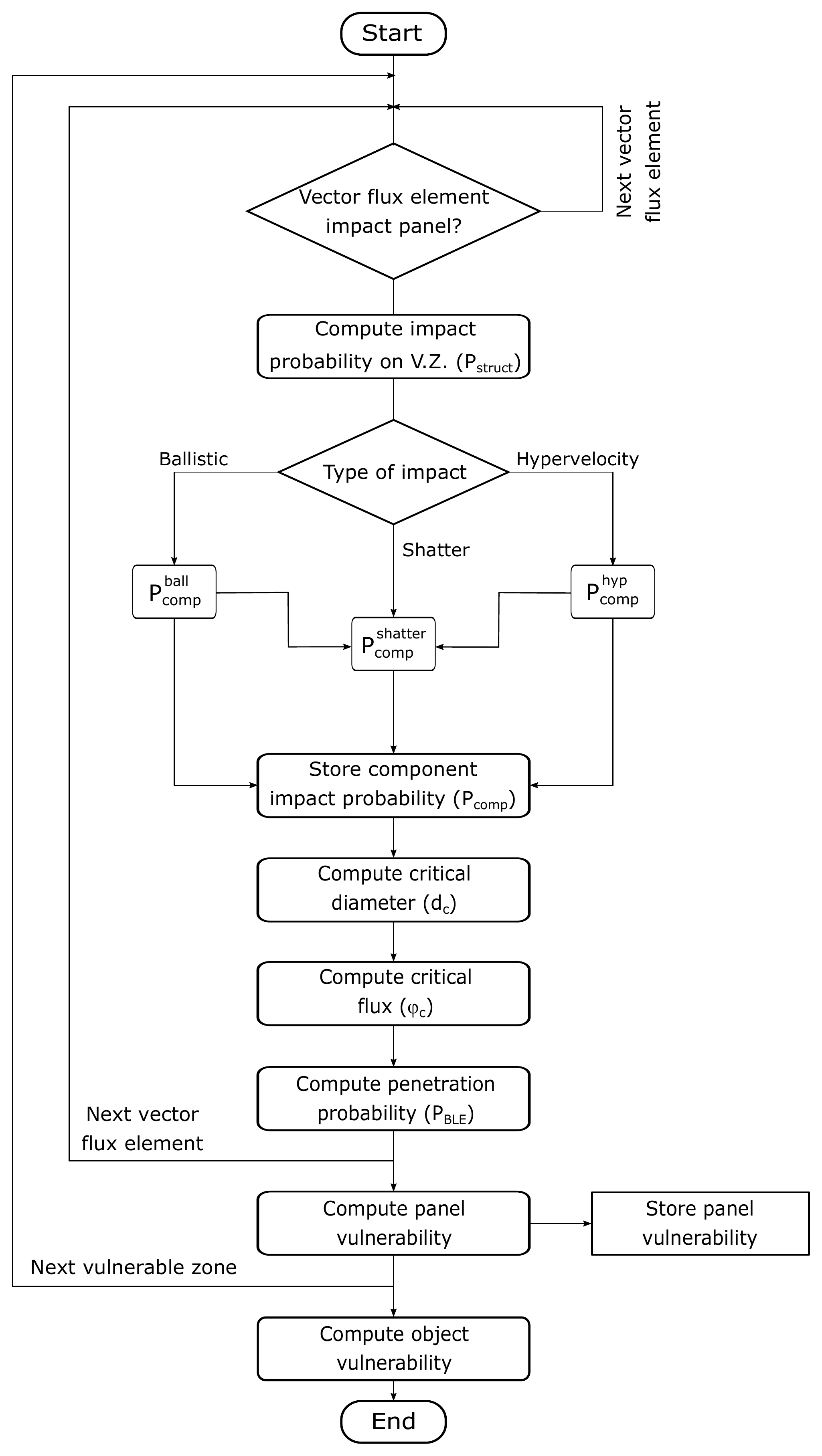}
	\caption{Flow diagram for the vulnerability computation of an internal component.}
	\label{fig:vuln_diagram}
\end{figure}

For a complete vulnerability assessment, it is also necessary to consider the effects on internal components. When a debris particle with sufficient size and velocity impacts the outer structure of the spacecraft, a secondary debris cloud is usually generated \citep{Putzar2006, Depczuk2003}. The particles belonging to this secondary cloud can impact the internal components; therefore, it is necessary to evaluate the probability that the secondary ejecta have to damage internal components. This can be achieved through a fully statistical procedure based on the concept of vulnerable zones \citep{Trisolini2018_AESCTE, Putzar2006, Welty2013}, where the vulnerability of an internal component ($V_{comp}$) can be expressed as the product of three different contributions as follows

\begin{equation} \label{eq:vulnerability}
V_{comp} = P_{struct} \cdot P_{cloud} \cdot P_{BLE}
\end{equation}

where $P_{struct}$ is the probability of space debris hitting the spacecraft external structure inside the vulnerable zone relative to the target component, $P_{cloud}$ is the probability that the secondary cloud ejecta will hit the component, and $P_{BLE}$ is the probability that the particles in this cloud perforate the component wall. \cref{fig:vuln_diagram} shows the schematics of the entire procedure.

\subsection{Impact propagation}
\label{subsec:impact}
\begin{figure}[htb!]
\centering
\includegraphics[width=0.6\textwidth]{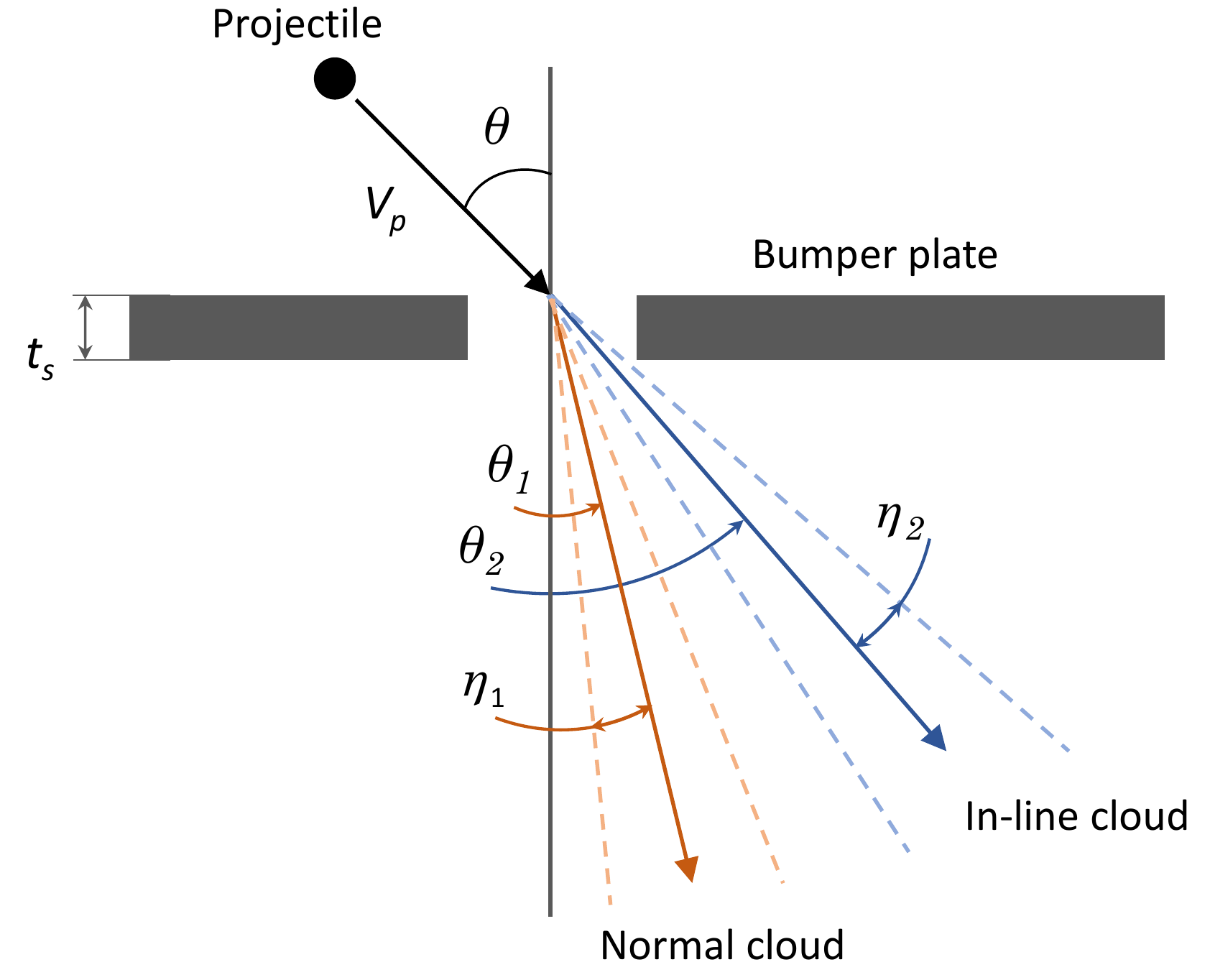}
	\caption{Geometry of secondary cloud ejecta.}
	\label{fig:impact_prop}
\end{figure}

Understanding how debris impacts propagate inside the spacecraft plays a crucial role in modelling and predicting the damage received by internal components due to secondary debris ejecta. Experimental campaigns on very high-speed impacts have shown a considerable difference between ballistic and hypervelocity behaviour. In the former case, the projectile stays almost intact after the impact and keeps propagating in a direction that almost coincides with the impact direction. In the latter case, instead, both the projectile and the impacted plate suffer a fragmentation, which produces two secondary clouds of debris (\cref{fig:impact_prop}) \citep{Schonberg2001, Depczuk2003}. One cloud exits almost perpendicularly to the impacted wall and is referred to as the \emph{normal debris cloud}. The second cloud instead follows more closely the direction of the projectile and is identified as the \emph{inline debris cloud}. It is assumed that the debris belonging to these clouds remains contained inside conic surfaces so that their behaviour can be modelled using only the characteristics of the cone (i.e. the direction of the cone axis and the spread angle of the cone).
\crefrange{eq:ejecta_1}{eq:ejecta_4} describes the relation between the characteristics of the impact (impact velocity, impact angle, particle diameter, and wall material) and geometry of the cones \citep{Schonberg2001, Depczuk2003}.
\begin{align}
\frac{\theta_1}{\theta} &= 0.471\cdot\Big( \frac{v_p}{C} \Big)^{-0.049}\cdot\Big( \frac{t_s}{d_p} \Big)^{-0.054} \cdot \cos{(\theta)}^{1.134} \label{eq:ejecta_1} \\
\frac{\theta_2}{\theta} &=  0.532\cdot\Big( \frac{v_p}{C} \Big)^{-0.086}\cdot\Big( \frac{t_s}{d_p} \Big)^{-0.478} \cdot \cos{(\theta)}^{0.586} \label{eq:ejecta_2} \\
\tan{\eta_1} &=  1.318 \cdot \Big( \frac{v_p}{C} \Big)^{0.907} \cdot \Big( \frac{t_s}{d_p} \Big)^{0.195} \cdot \cos{(\theta)}^{0.394} \label{eq:ejecta_3} \\
\tan{\eta_2} &=  1.556 \cdot \Big( \frac{v_p}{C} \Big)^{1.096} \cdot \Big( \frac{t_s}{d_p} \Big)^{0.345} \cdot \cos{(\theta)}^{0.738} \label{eq:ejecta_4}
\end{align}
where $\theta$ is the impact angle, $\theta_1$ is the deflection angle of the normal debris cloud, $\theta_2$ is the deflection angle of the inline debris cloud, $\eta_1$ is the half-cone angle of the normal cloud, $\eta_2$ is the half-cone angle of the inline cloud. $t_s$ is the impacted plane thickness, \textit{C} is the speed of sound of the plate material, $d_p$ is the particle diameter, and $v_p$ is the relative particle impact velocity.
\subsection{Vulnerable zones}
\label{subsec:vulnerable_zone}
\begin{figure}[htb!]
\centering
\includegraphics[width=0.6\textwidth]{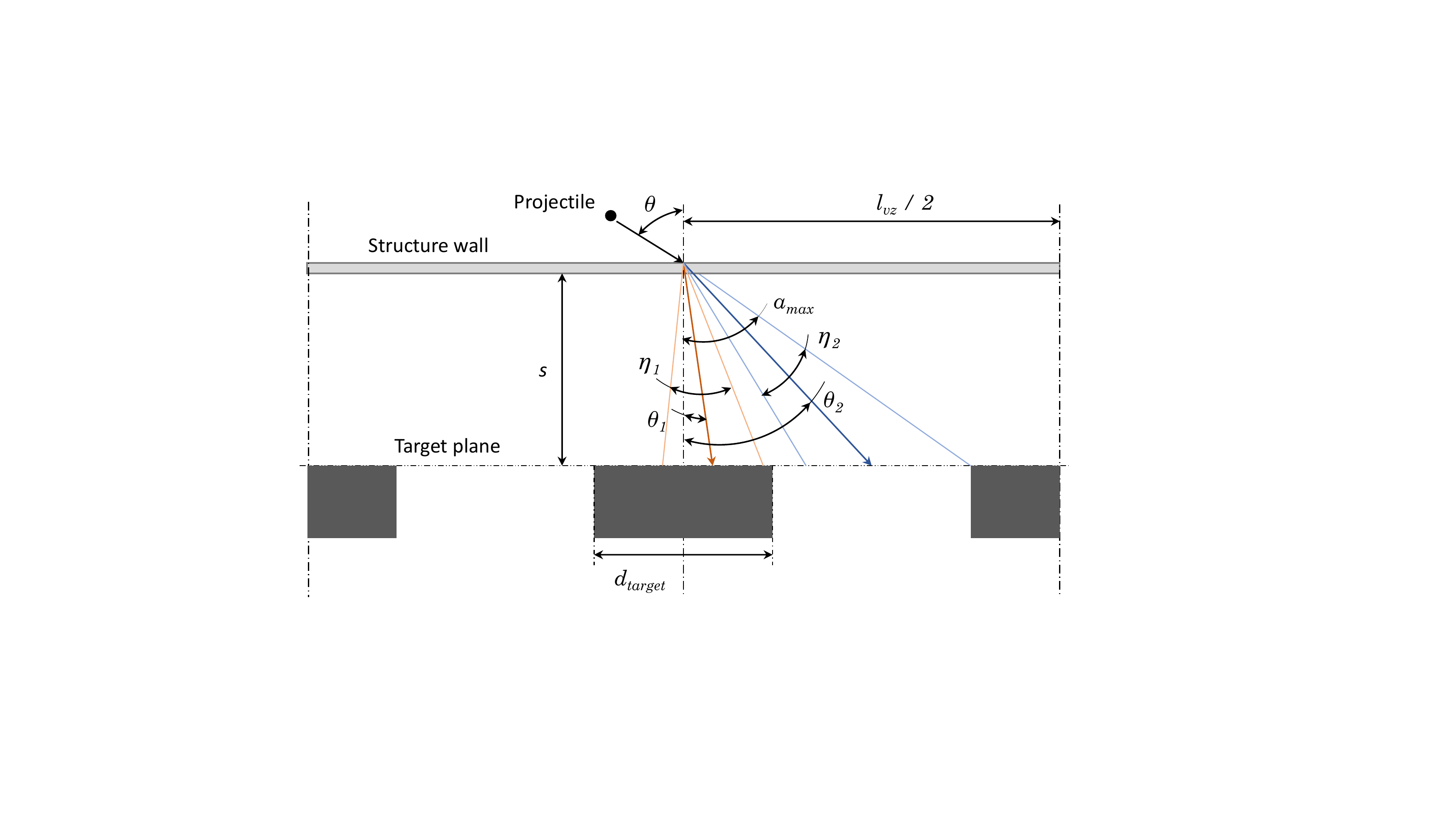}
	\caption{Representation of the lateral extension of the vulnerable zone.}
	\label{fig:vzone}
\end{figure}

\begin{figure}[htb!]
\centering
\includegraphics[width=0.55\textwidth]{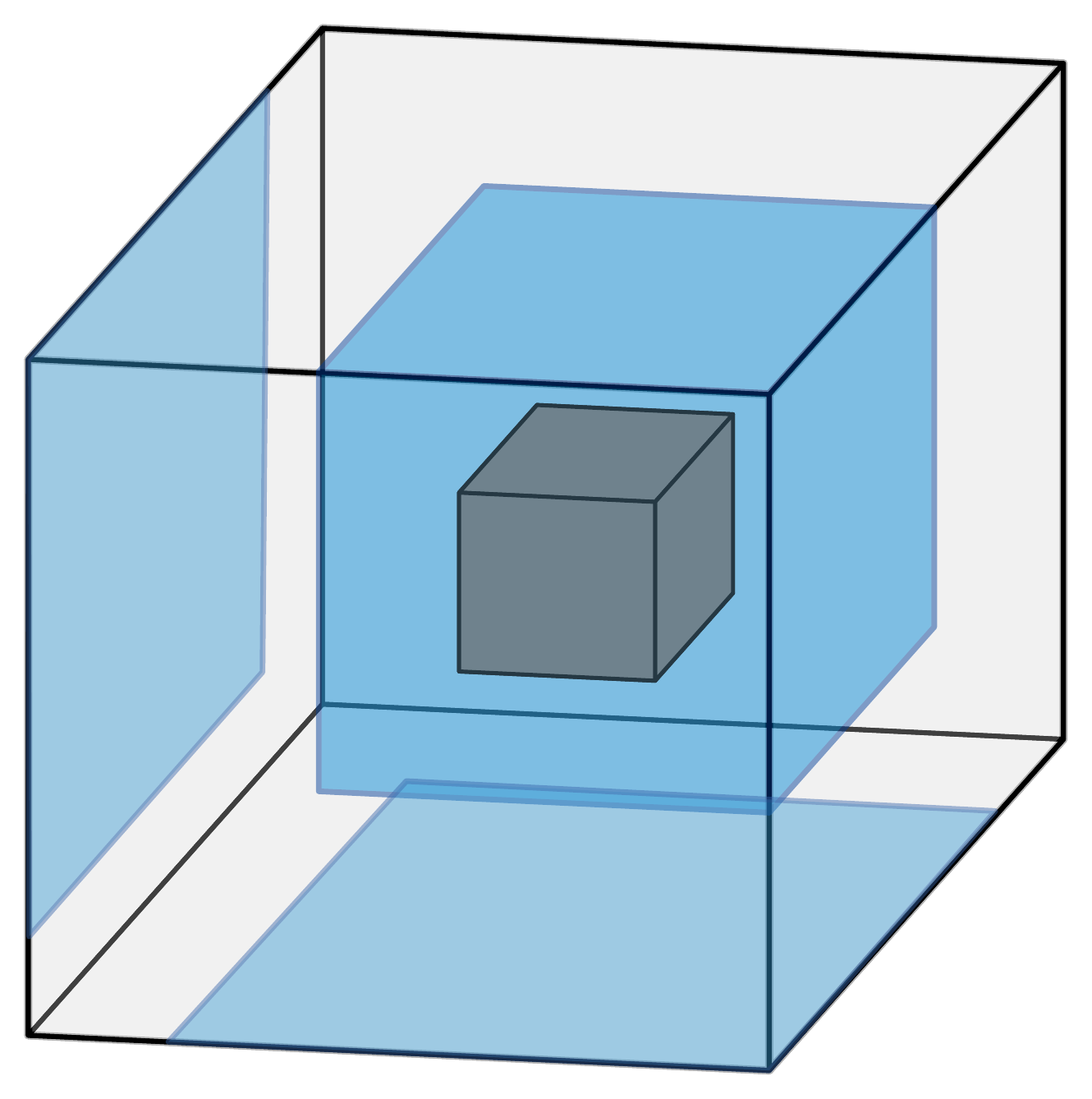}
	\caption{Vulnerable zones of a box projected onto the faces of a cubic structure (the vulnerable zone of the back face is omitted for clarity).}
	\label{fig:vzone_draw}
\end{figure}

The vulnerable zone \citep{Putzar2006} is defined as the area on the external structure of the spacecraft that, if impacted, can lead to an impact also on the considered component (\cref{fig:vzone}). Any impact of a particle onto this area generates fragments that may hit the component in question, with a probability that depends on the geometry of the impact (secondary debris ejecta characteristics, stand-off distance, and shielding from other components). The lateral extent of the vulnerable zone (\textit{$l_{vz}$}) is expressed as
\begin{equation} \label{eq:vz}
l_{vz} = 2 \cdot \Big( s \cdot \tan{\alpha_{max}} + \frac{1}{2} \cdot \big( d_{target} + d_{p, max} \big) \Big)
\end{equation}
where \textit{s} is the spacing between the structure wall and the component front face (stand-off distance), $d_{target}$ is the lateral extent of the target component (the component of which we are evaluating the vulnerability), $d_{p, max}$ is the maximum projectile diameter, and $\alpha_{max}$ is the maximum ejection angle, which has a value of 63.15 degrees \citep{Putzar2006}. The parameter $d_{p, max}$ instead represents a user-defined value for the extent of the particle diameter, which takes into account the contribution of the particle to the impact probability. Suggested values for $d_{p, max}$ are 10 mm for vulnerable components and 20 mm for components with higher impact resistance \citep{Putzar2006}. \cref{fig:vzone_draw} shows a visual example of the vulnerable zones for a box-shaped component inside a cubic structure. The cyan regions represent the projection of the vulnerable zones extensions onto the external structure.
\subsection{Penetration probability of internal components}
\label{subsec:Ppen_internal_comp}
As previously introduced in \cref{eq:vulnerability}, the computation of the penetration probability on an internal component is subdivided into the computation of three different contributions. The first contribution, $P_{struct}^{ij}$, is calculated following the procedure outlined in \cref{subsec:prob_assessment}. However, in this case, the surface to be considered is the one delimited by the vulnerable zone (\cref{fig:vzone_draw}). The resulting expression is
\begin{equation} \label{eq:p_struct}
P_{struct}^{ij} = 1-\exp{\big( -\varphi_i \cdot S_{vz}^{ij} \cdot \Delta T} \big)
\end{equation}
where $P_{struct}^{ij}$ is the impact probability on the \textit{j}-th vulnerable zone by the \textit{i}-th vector flux element. $\varphi_i$ is the flux relative to the \textit{i}-th vector flux element, $S_{vz}^{ij}$  is the projected area of the \textit{j}-th vulnerable zone relative to the target component along the direction of the \textit{i}-th \emph{vector flux element}; and $\Delta T$ is the mission time in years.
When a particle hits the vulnerable zone, it does not necessarily mean that the downrange fragments will damage the target component.
The second term in \cref{eq:vulnerability} takes this into account. To properly compute this contribution, it is necessary to consider the three different impact regimes (\emph{ballistic}, \emph{hypervelocity}, and \emph{shutter}) separately; in fact, they lead to different geometries for the impact ejecta and require different expressions for the assessment of the critical diameter (\cref{subsec:ble}). In case of an hypervelocity impact, the probability to impact the target component is computed as the ratio between the extent of the cloud ejecta at the target component plane and the extent of the vulnerable zone of the component \citep{Putzar2006}. The resulting expression is given by
\begin{equation} \label{eq:p_comp_hyp}
P_{cloud, h}^{ij} = \frac{d_{ejecta}^{ij}}{l_{vz,j}}
\end{equation}
where $P_{cloud,h}^{ij}$ is the probability that the cloud relative to the \textit{i}-th vector flux element, which has already impacted the \textit{j}-th vulnerable zone, will hit the target component, $l{vz, j}$ is the extent of the \textit{j}-th vulnerable zone in the target plane, and $d_{ejecta}^{ij}$ is the extent of the debris ejecta at the target plane relative to the \textit{j}-th vulnerable zone (\cref{fig:vzone}), which is expressed as 
\begin{equation} \label{eq:d_ejecta}
d_{ejecta}^{ij} = 2 \cdot \Big( \tan{\alpha\big( \theta_{ij} \big)} \cdot s_j + 1/2 \cdot d_{target,j} \Big)
\end{equation}
where $s_j$ is the stand-off distance between the component and the external wall to which the \textit{j}-th vulnerable zone is associated, $d_{target,j}$ is the extent of the target component along the direction of the \textit{j}-th vulnerable zone, and $\alpha(\theta_{ij})$ is the ejection angle associated with the \textit{i}-th vector flux element impacting on the \textit{j}-th vulnerable zone, and can be computed with the following expression
\begin{equation} \label{eq:ejection_angle}
\alpha(\theta) = \theta_2 + \frac{\eta_2}{2}
\end{equation}
which is obtained simplifying \crefrange{eq:ejecta_1}{eq:ejecta_4}, assuming that the ejection and spread angles are only a function of the impact angle $\theta$ and that all the other parameters can be absorbed by a constant factor \citep{Putzar2006}. 
In case of an impact in the ballistic regime, as no fragmentation occurs, only the size of the projectile needs to be taken into account.
\begin{equation} \label{eq:p_comp_ball}
P_{cloud, b}^{ij} = \frac{d_{target}^{ij} + d_p^{ij}}{l_{vz,j}}
\end{equation}
where $d_p^{ij}$   is the particle diameter relative to the \textit{i}-th vector flux element impacting on the \textit{j}-th vulnerable zone. This value is the most probable particle diameter for the \textit{i}-th vector flux element (\cref{subsubsec:vfe}) and is extracted from the debris flux distributions from \gls{master}-2009. For the scatter regime, a linear interpolation (\cref{eq:p_comp_shatter}) between the hypervelocity regime (\cref{eq:p_comp_hyp}) and the ballistic regime (\cref{eq:p_comp_ball}) is adopted.
\begin{equation} \label{eq:p_comp_shatter}
P_{cloud,s}^{ij} = P_{cloud, b}^{ij} + \bigg( \frac{v_{p}^{ij} - V_{LV}}{V_{HV} - V_{LV}} \bigg) \cdot P_{cloud, h}^{ij}
\end{equation}
The value of $P_{cloud}$ used in \cref{eq:vulnerability} is then selected as follows
\begin{equation} \label{eq:Pcloud_cases}
P_{cloud}^{ij} =
\begin{cases}
P_{cloud,b}^{ij} &\text{if $v_p \leq V_{LV}\cdot\cos{\theta}$ }  \\[6pt]
P_{cloud,h}^{ij} &\text{if $v_p \geq V_{HV}\cdot\cos{\theta}$ }  \\[6pt]
P_{cloud,s}^{ij} &\text{otherwise}
\end{cases}
\end{equation}
Finally, the last contribution in \cref{eq:vulnerability} is related to the computation of the penetration probability on the front face of the target component. Similarly, the contribution can be expressed as
\begin{equation} \label{eq:p_ble}
P_{BLE}^{ij} = 1- \exp{\big( -\varphi_{c,i} \cdot S_{vz,ij} \cdot \Delta T} \big)
\end{equation}
where $P_{BLE}^{ij}$  is the penetration probability for the \textit{j}-th vector flux element on the component associated with the \textit{i}-th vulnerable zone, and $\varphi_{c,i}$ is the critical flux that is the flux associated to the value of the critical diameter computed with \glspl{ble}. As specified in Section \ref{subsec:ble}, we adopted the \gls{srl} \gls{ble} to compute the critical diameter relative to the target component. It is always assumed that the last wall of the shielding configuration corresponds to the face of the target component, while the other walls are representative of the outer structure. The overall vulnerability of an internal component can then be expressed as
\begin{equation} \label{eq:Vp}
V_{comp} = 1 - \prod_{j=1}^{N_{panels}} \Bigg( \prod_{i=1}^{N_{fluxes}} \Big( 1 - P_{struct}^{ij} \cdot P_{cloud}^{ij} \cdot P_{BLE}^{ij} \Big) \Bigg)
\end{equation}
\section{Mutual shielding methodology}
\label{subsec:mutual_shielding}
\begin{figure}[htb!]
\centering
\includegraphics[height=7cm, keepaspectratio]{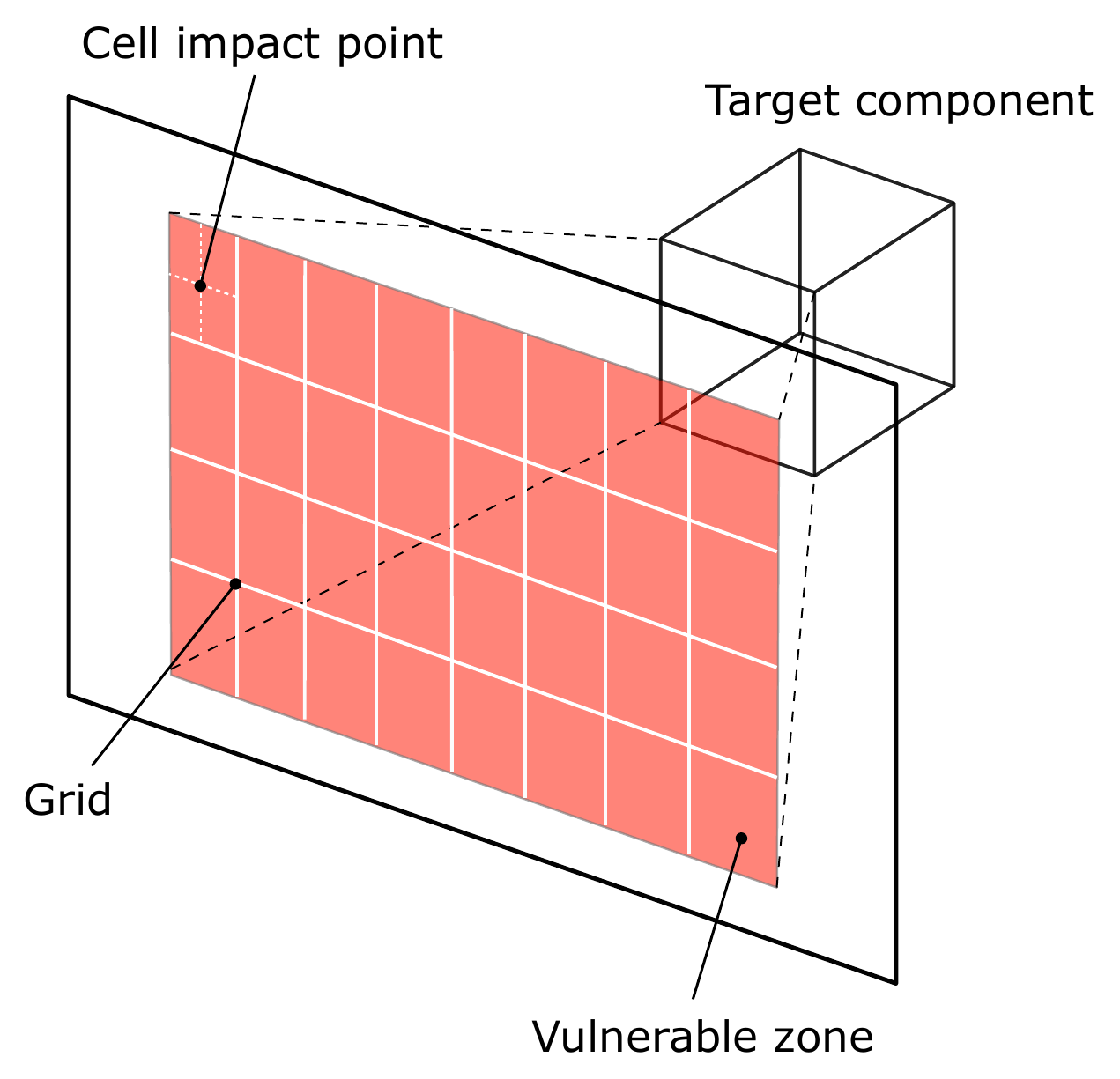}
	\caption{Representation of the vulnerable zone grid and of the cell impact locations.}
	\label{fig:vz_grid}
\end{figure}

The vulnerable zone methodology as described in \cref{sec:vulnerability} lacks the capability of considering the mutual shielding between components. In fact, \cref{eq:pc_hyp,eq:pc_ball} do not take into account the contribution of possible shadowing between the components, where additional shielding is provided, preventing part if not all the secondary debris ejecta from impacting them. This is especially important considering the directional nature of the space debris fluxes. For example, impacts coming from the RAM direction are more dangerous than from other directions because of a higher relative velocity.
As the vulnerable zone approach has its simplicity and the possibility to avoid time-consuming Monte Carlo simulations among its advantages, the idea is to extend this approach by integrating a methodology that allows the evaluation of the mutual shielding while still maintaining the advantages of the original approach. Among the three different contributions to the vulnerability (\cref{eq:vulnerability}), the second term (associated with the impact probability of the ejecta on the component) is the one connected to the mutual shielding. The idea is to consider the interaction between the debris cone developed after the impact and the shielding elements interposed between the outer faces and the component in a purely geometrical way. To do so, we first consider the nature of the impacts on the spacecraft structure. In the standard vulnerable zone approach, a particle can impact anywhere on the vulnerable zone (no impact location is sampled or specified). The lack of a precise impact location results in an issue connected to the assessment of the mutual shielding. In fact, the shadowing between components depends on the impact location and on the characteristic of the debris produced after the impact. In case of a hypervelocity impact, a cone of debris is generated; depending on the cone axis and aperture angle, and on the impact location, different areas of the target component may be visible and different portions of the debris cone can be shielded by the interposing components. To overcome this issue it was decided to discretize the vulnerable zone using a grid (\cref{fig:vz_grid}). As there must not be a preferred impact location, for each vector flux element, an impact is simulated assuming the centre of each cell in the grid to represent the impact location. At this point, for each impact location, the resulting debris cone is generated and its interaction with the target and the shielding components is evaluated. 
The overall effect of a vector flux element impacting the considered vulnerable zone is then obtained averaging the contribution of each cell in the grid. The procedure is then repeated for each vector flux element and each vulnerable zone. 
In the case of a ballistic impact, the procedure must change. In this case, the projectile passes through in the same direction of the impact vector, and no secondary debris is produced.
\subsection{Hypervelocity regime}
\label{subsec:hypervelocity}
\begin{figure}[htb!]
\centering
\includegraphics[height=5.5cm, keepaspectratio]{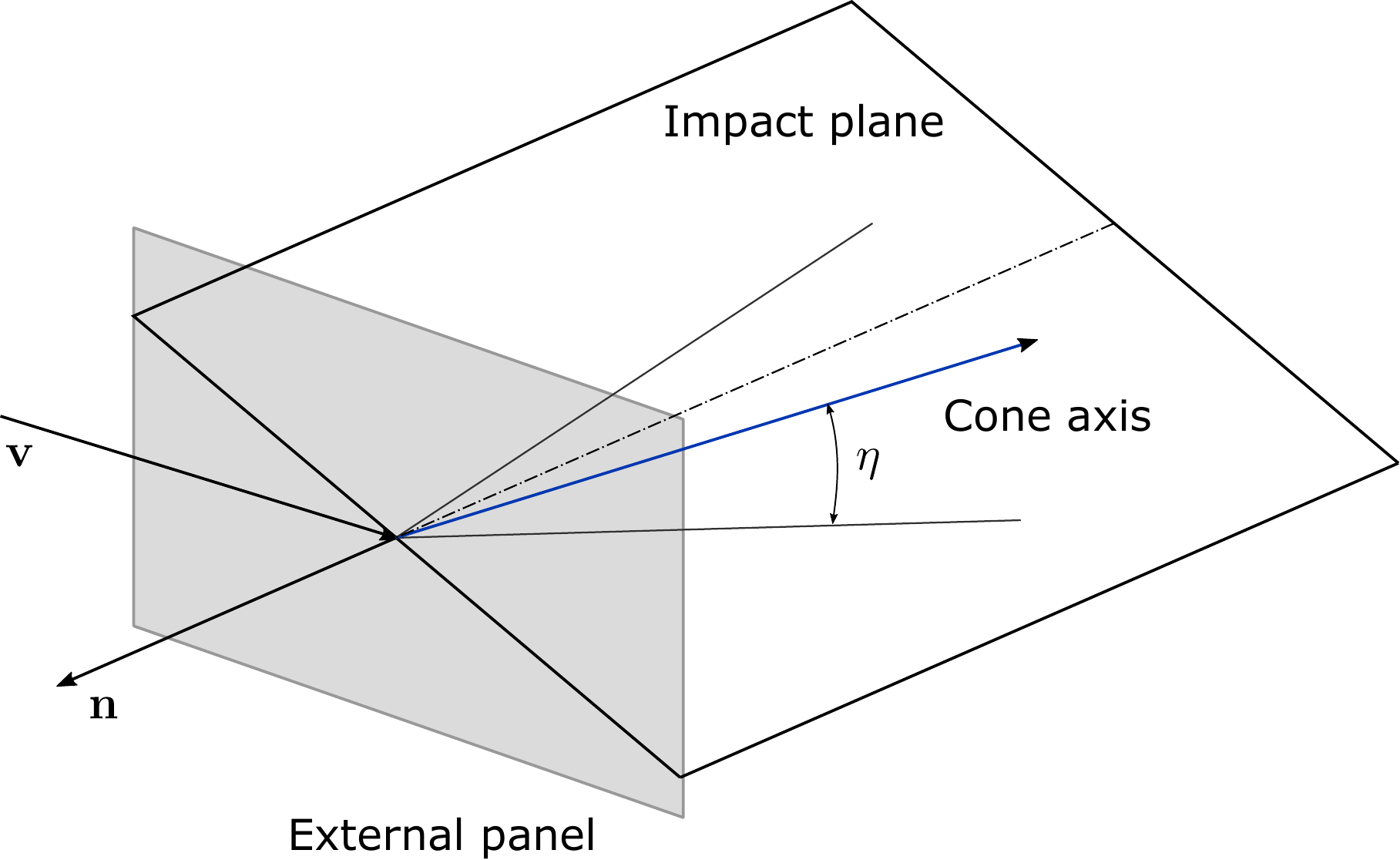}
	\caption{Representation of the cone aperture angle in the impact plane.}
	\label{fig:impact_plane}
\end{figure}

\begin{figure}[htb!]
\centering
\includegraphics[height=7cm, keepaspectratio]{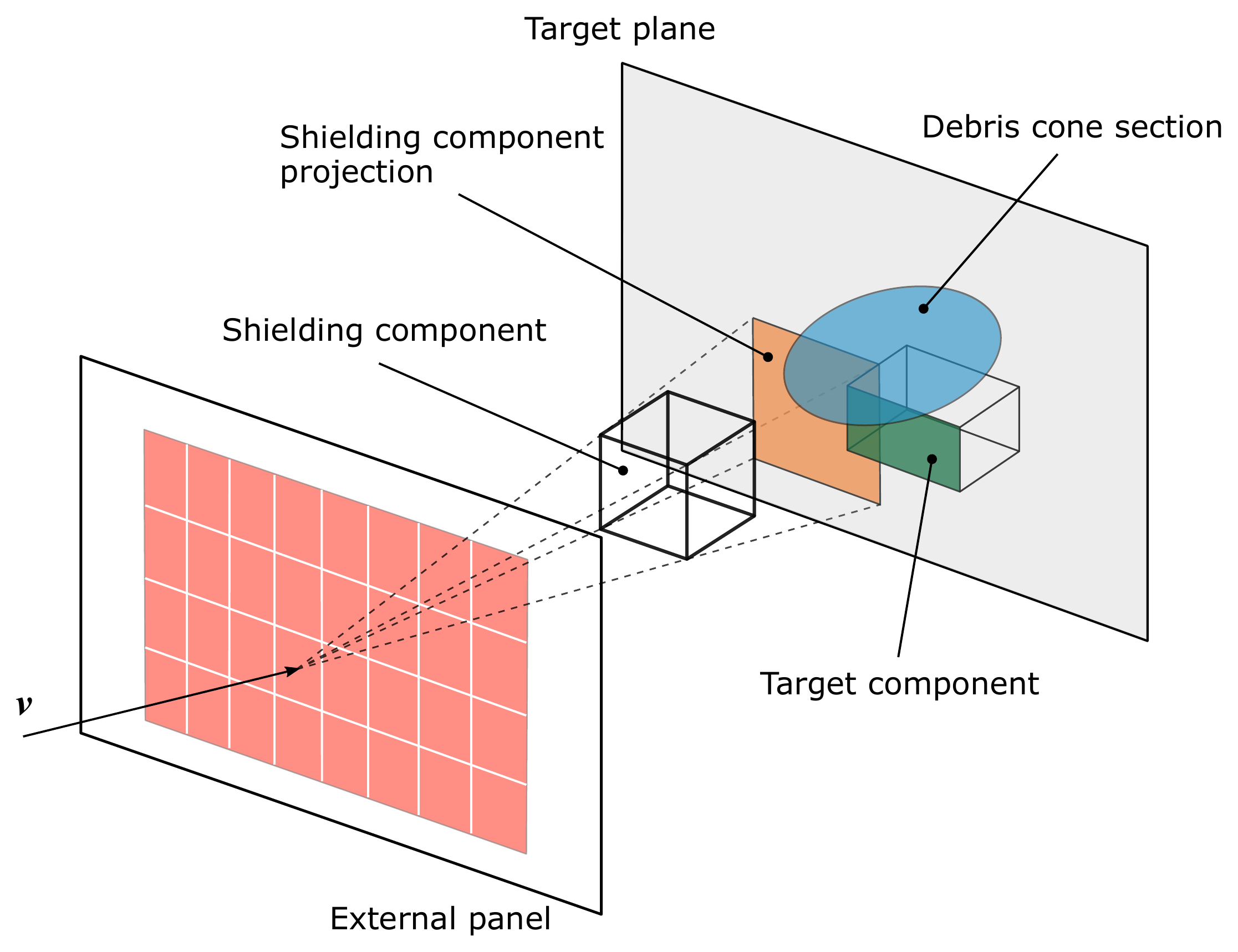}
	\caption{Mutual shielding in the hypervelocity regime.}
	\label{fig:hyp_impact}
\end{figure}

After a hypervelocity impact, the impacting particle is destroyed together with the area of the panel subject to the impact, and two secondary clouds of fragments are generated (\cref{subsec:impact}). In the standard vulnerable zone formulation (\cref{subsec:vulnerable_zone}), the extent of the vulnerable zone is defined using a single conical shape (\cref{eq:vz}). To maintain this approach, the two secondary clouds are merged into a single conical shape. The characteristics of the cone (i.e. its axis and aperture angle) are determined using \crefrange{eq:ejecta_1}{eq:ejecta_4}. The cone half-aperture angle ($\eta$) can be computed as follows:
\begin{equation} \label{eq:cone_angle}
\eta = \frac{1}{2} \cdot (\theta_2 - \theta_1) + \frac{1}{4} \cdot (\eta_1 + \eta_2)
\end{equation}
The axis of the cone is the bisector of the aperture angle and belongs to the impact plane, which is the plane containing the impact vector flux element and the normal to the impacted face (\cref{fig:impact_plane}). It is oriented of an angle equal to $\eta - \theta_1$ with respect to the vector normal to the external plane. The impact plane defines the interaction between the debris cone and the target plane, which is used to define the debris cone section at the impact plane (\cref{fig:hyp_impact}). In fact, it is used to determine the position and orientation of the debris cone section: the intersection of the axis of the cone and the target plane is the centre of ellipse and the line belonging to the impact plane and perpendicular to the cone axis will be its semi-major axis. The cone obtained is then used to determine the impact probability on the target component. To do so, the interaction of the debris cone with the target object and the shielding components must be evaluated.

The standard approach computes the impact probability by adding the extent of the target section to the extent of the debris cone in the target plane (\cref{eq:p_comp_hyp}). This mechanism is extended here, and the mutual shielding is accounted for in both the debris cone and of the target section. For the debris cone, a perspective projection of the components onto the target plane is carried out and the intersection of these projections with the section of the cone at the target plane is performed (\cref{fig:hyp_impact}).

The area used in the computation is then the residual area of the cone after the shielding component projections have been subtracted. This area is referred to as the \emph{available cone area} $A_{c,av}$. For each vector flux element, this area is evaluated over all the grid cells subdividing the vulnerable zone.

\begin{equation} \label{eq:av_cone}
A_{c,av}^{k} = A_c^k \neg \big( A_{s,1}^{k},...,A_{s,n}^{k},...,A_{s,N_s}^{k} \big), \qquad k \in \{ 1,...,N_{cell} \};
\end{equation}

where $A_{c}^{k}$ is the intersection ($\neg$) between the \textit{k}-th debris cone with the target plane, $A_{s,n}^{k}$ is the perspective projection of the \textit{n}-th shielding component onto the target plane with respect to the \textit{k}-th grid cell, $N_s$ is the number of shielding components between the target component and the \textit{j}-th vulnerable zone, and $N_{cell}$ is the number of cells subdividing the vulnerable zone. The procedure is repeated for each grid cell and the result averaged in order to obtain the \emph{average available cone area} relative to the \textit{i}-th vector flux element impacting the \textit{j}-th vulnerable zone ($\bar{A} _{c, av}^{ij}$)

\begin{equation} \label{eq:av_cone_ji}
\bar{A}_{c,av}^{ij} = \frac{1}{N_{cell}} \cdot \sum_{k=1}^{N_{cell}} A_{c,av}^{k}
\end{equation}

\begin{figure}[!htb]
\centering
\includegraphics[height=7cm, keepaspectratio]{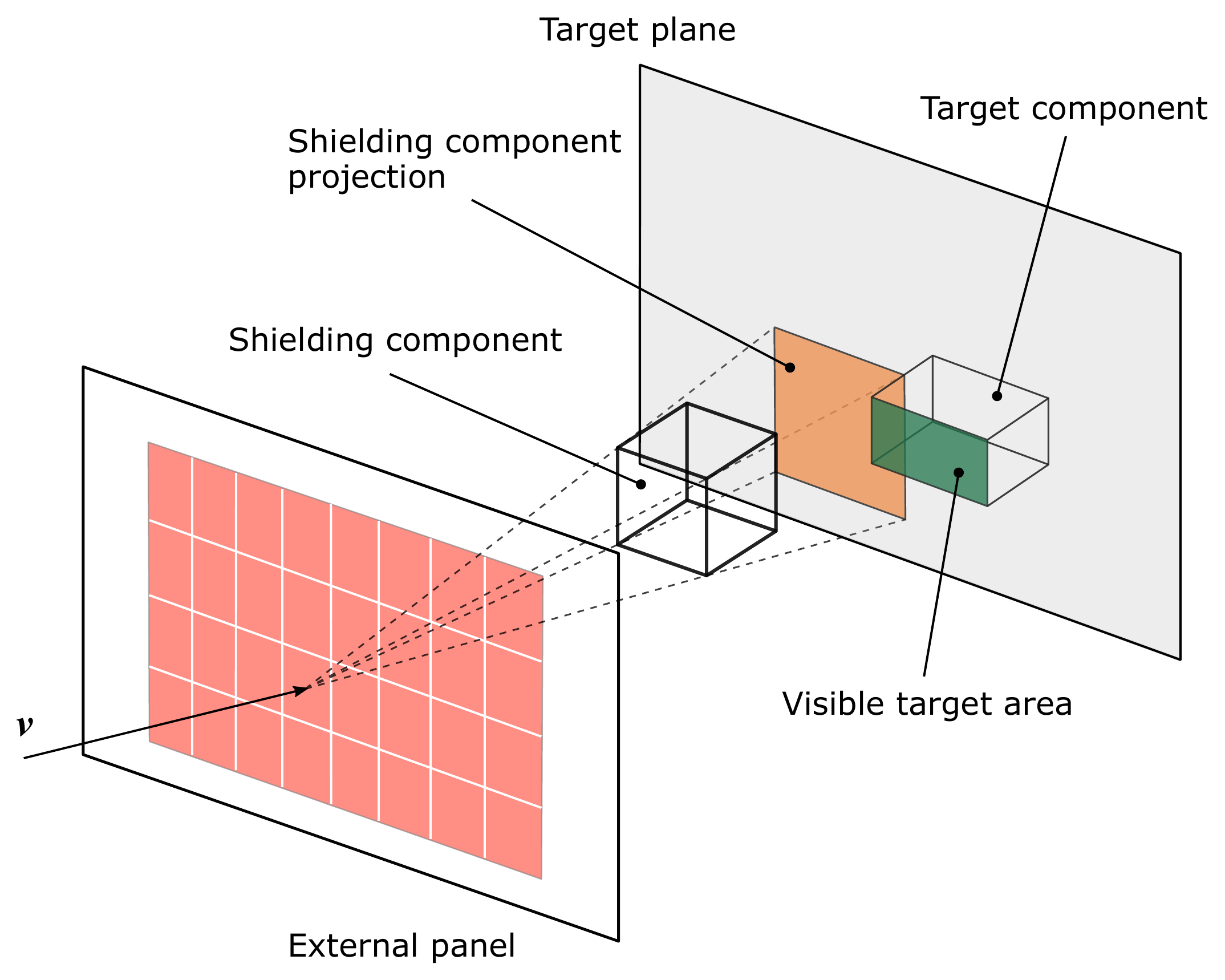}
	\caption{Perspective projection of a shielding component onto the target plane.}
	\label{fig:ball_impact}
\end{figure}

As the projection of the components and the cone section at the target plane can both exceed the limits of the spacecraft envelope, all the computed areas are cropped with respect to the limits of the target plane, which is limited by the external structure of the spacecraft. All the boolean operations such as intersection and difference between the projected areas have been performed using the Python package Shapely \citep{Gillies2007}.
The second contribution in the vulnerable zone equation is the target component length (\cref{eq:vz}). In this extension of the methodology, the \emph{visible target area} is computed by performing a Boolean difference between the target section and the perspective projection of the shielding components onto the target plane (\cref{fig:ball_impact}). Again, the operation is repeated over each grid cell, and the average over the overall grid is carried out to evaluate the \emph{average target visible area} ($A_{t, av}^{ij}$) as follows:
\begin{equation} \label{eq:av_target}
\bar{A}_{t,av}^{ij} = \frac{1}{N_{cell}} \cdot \sum_{k=1}^{N_{cell}} \Big[ A_t \neg \big( A_{s,1}^{k},...,A_{s,n}^{k},...,A_{s,N_s}^{k} \big) \big]
\end{equation}
The impact probability associated to the \textit{i}-th vector flux element impacting on the \textit{j}-th vulnerable zone for the hypervelocity regime is then obtained with the following expression
\begin{equation} \label{eq:pc_hyp}
P_{cloud,h}^{ij} = \frac{\bar{A}_{t,av}^{ij}+\bar{A}_{c,av}^{ij}}{A_{vz,j}}
\end{equation}
where $A_{vz,j}$ is the area of the j-th vulnerable zone.
\subsection{Ballistic regime}
\label{subsec:ballistic}
\begin{figure}[!htb]
\centering
\includegraphics[height=5.5cm, keepaspectratio]{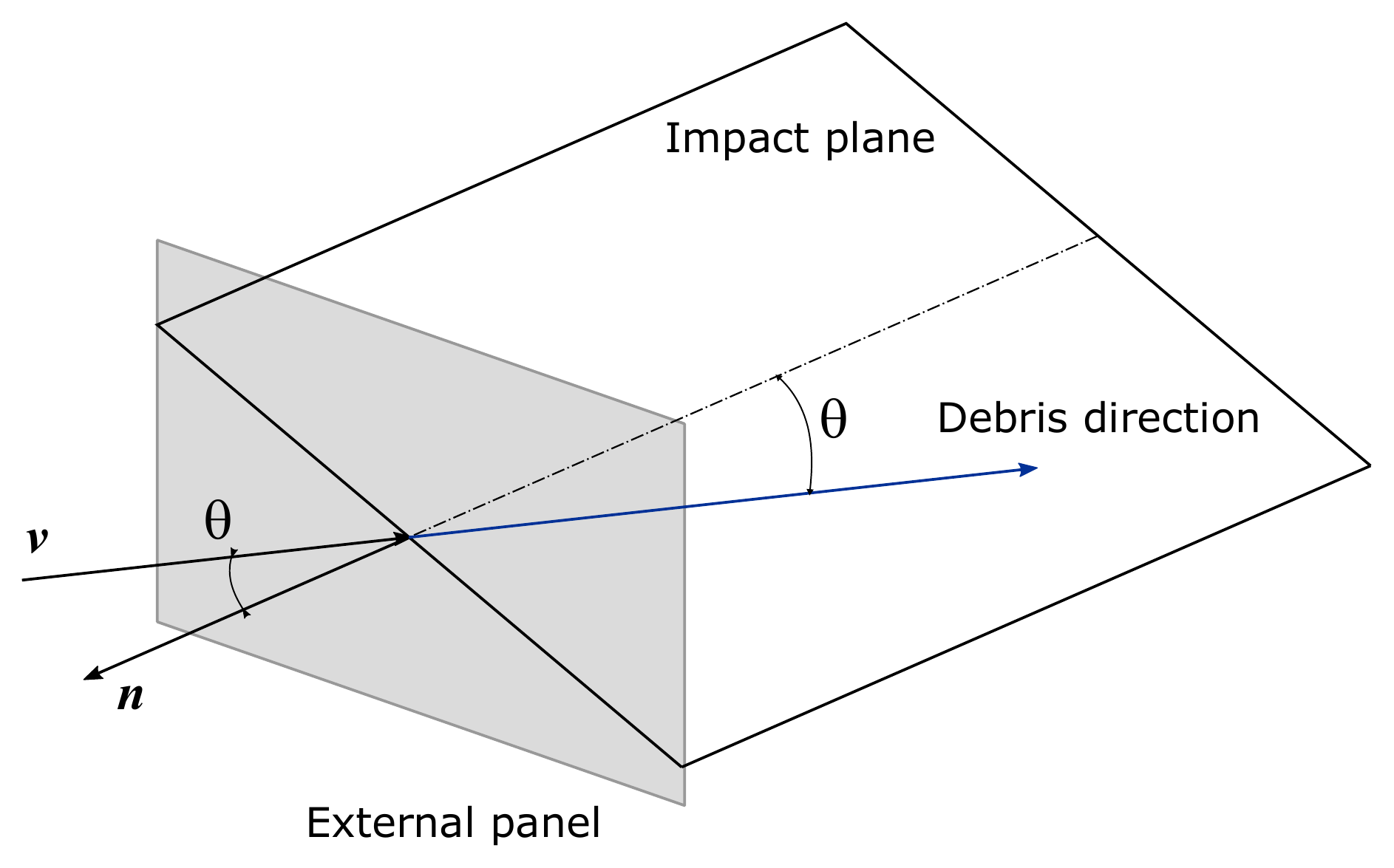}
	\caption{Trajectory of the vector flux elements after a ballistic impact.}
	\label{fig:ball_imp_plane}
\end{figure}

In case of a ballistic impact, the projectile passes through the panel without being destroyed. Therefore, no ejecta are produced and the consequence of the impact cannot be schematized with a debris cone. Instead, for each vector flux element and impact point on the vulnerable zone grid, a line is generated with a vertex on the centre of the cell and direction equivalent to the one of the vector flux element (\cref{fig:ball_imp_plane}).
This line represents the trajectory of the debris after the impact. The interaction of this line with the target component and the shielding components is then evaluated. Again, this is to maintain the analogy with the standard formulation. The variables that need to be considered are the particle size and the target area. It is assumed that the particle is not affected during the impact, maintaining its shape (spherical) and dimension. The equivalence is obtained by using the cross-section of the particle (\cref{eq:particle_section}).
\begin{equation} \label{eq:particle_section}
A_{p,i} = 1/4 \cdot \pi \cdot d_{p,i}^2
\end{equation}
where $d_{p,i}$ is the sample particle diameter associated with the \textit{i}-th vector flux element. For the target section, the same procedure used in the hypervelocity case is adopted and the computation of the \emph{average target visible area} is performed using \cref{eq:av_target}. Finally, the impact probability in the ballistic case is given as
\begin{equation} \label{eq:pc_ball}
P_{cloud,b}^{ij} = \frac{\bar{A}_{t,av}^{ij}+A_{p,i}}{A_{vz,j}}
\end{equation}
\cref{eq:pc_hyp,eq:pc_ball} can be used inside \cref{eq:Vp} to compute the impact probability of an internal component.

\subsection{Equivalence with standard vulnerable zone formulation}
\label{subsec:linear_impact}
\cref{subsec:hypervelocity,subsec:ballistic} describe the methodology for assessing the contribution of shielding components through a procedure that relies on the use of areas, as this is the more natural way of considering such contribution. However, the standard vulnerable zone formulation adopts an approach that uses the linear extension of the vulnerable zone and of the target component (\cref{subsec:vulnerable_zone}). As the developed methodology aims at extending the standard formulation, \cref{eq:pc_hyp,eq:pc_ball} have been converted into a linearised version as follows

\begin{align}
P_{comp,h}^{\,ij} &= \frac{\bar{d}_{t,av}^{\,ij}+\bar{d}_{c,av}^{\,ij}}{l_{vz,j}} \label{eq:pc_hyp_lin} \\
P_{comp,b}^{\,ij} &= \frac{\bar{d}_{t,av}^{\,ij}+d_{p,i}}{l_{vz,j}} \label{eq:pc_ball_lin}
\end{align}

where $\bar{d}_{t,av}$ and $\bar{d}_{c,av}$ maintain the same meaning. The conversions from the areas to the equivalent length is carried out using the following simplified expression

\begin{equation} \label{eq:d_eq}
d = 2 \cdot \sqrt{A/\pi}
\end{equation}

For completeness, both the methodology using the areas (\cref{eq:pc_hyp,eq:pc_ball}) and the one using the equivalent linear extent (\cref{eq:pc_hyp_lin,eq:pc_ball_lin}) have been tested and compared with state-of-the-art software packages. The results are presented in \cref{sec:comparison}.

\subsection{Correction factor methodology}
\label{subsec:correction_factor}
The previously described methodology captures the geometrical properties of the impact phenomenon and its propagation inside the spacecraft structure; however, it still requires a computational time that is not applicable to a multi-objective optimisation framework (around 6 seconds for a configuration with two internal components for a Python implementation). As the presented model is meant to be used inside a multi-objective optimisation framework, its execution speed is of paramount importance. Consequently, a simplified methodology has be developed, which still takes into account the mutual shielding, while reducing the large number of expensive geometrical operations needed for the complete procedure of \cref{subsec:hypervelocity,subsec:ballistic}. The method described in this section consists in applying a correction factor whenever one or more components cover the target component. This approach maintains the simplicity and the short computational time of the standard one while providing a consistent way to account for the reduced impact probability caused by the mutual shielding. Given the different nature of the impacts in the hypervelocity and in the ballistic regime, two correction factors are used.
In the case of the hypervelocity regime, it computes the portion of the vulnerable zone that can actually lead to an impact onto the target component. This is the part of the vulnerable zone that is not covered by the perspective projections of the shielding components. As these projections depend on the impact point, even in this case the vulnerable zone is subdivided into a grid and each grid cell centre is an impact point for the \emph{vector flux element}. The procedure is analogous to the one described in \cref{subsec:hypervelocity}; however, in this case, the perspective projections of the shielding components onto the target plane (\cref{fig:hyp_impact}) are directly subtracted from the vulnerable zone area. This methodology avoids generating the cone of the secondary debris ejecta and performing its intersection with the target plane that is the most computationally expensive part of the procedure. \cref{eq:av_cf} gives the expression for the \emph{available vulnerable zone area} $\bar{A}_{vz,av}^{ij}$ relative to the \textit{j}-th vulnerable zone and the can be computed again averaging over the contributions computed for each grid cell.

\begin{equation} \label{eq:av_cf}
\bar{A}_{vz,av}^{ij} = \frac{1}{N_{cell}} \cdot \sum_{k=1}^{N_{cell}} \Big[ A_{vz}^j \neg \big( A_{s,1}^{k},...,A_{s,n}^{k},...,A_{s,N_s}^{k} \big) \big]
\end{equation}

The expression for the hypervelocity correction factor is then

\begin{equation} \label{eq:cf_hyp}
C\!F_{h}^{ij} = \frac{\bar{A}_{vz,av}^{\;ij}}{A_{vz,j}}
\end{equation}

A value of 1 of the correction factor corresponds to no shielding, while a value of 0 indicates that the target component is not visible by the impactor. In the case of the ballistic regime, the same approach cannot be used, as there is no ejecta generation. Looking at \cref{eq:p_comp_ball}, the impact probability in the ballistic regime depends only on the size of the particle and on the extent of the target object. As the dimension of the particle cannot change, only the extent of the target can be changed in order to correct the impact probability. In the case of a ballistic impact, a corrected target extent is used, which can be referred to as the \emph{visible target area}. Using an approach similar to the hypervelocity case, the section of the shielding components is projected onto the target plane (\cref{fig:ball_impact}). At this point, if these projections intersect the target component, they are subtracted from it using Boolean operations. The procedure is repeated over each grid cell and averaged as follows:

\begin{equation} \label{eq:at_cf}
\bar{A}_{t,av}^{ij} = \frac{1}{N_{cell}} \cdot \sum_{k=1}^{N_{cell}} \Big[ A_{t,j} \neg \big( A_{s,1}^{k},...,A_{s,n}^{k},...,A_{s,N_s}^{k} \big) \big]
\end{equation}

where $\bar{A}_{t,av}^{ij}$ is the \emph{average visible target area} associated to the \textit{j}-th vulnerable zone and the \textit{i}-th \emph{vector flux element}, $A_{t,j}$ is the target component section relative to the \textit{j}-th vulnerable zone, and $A_{s, n}^k$  is the perspective projection of the \textit{n}-th shielding component onto the target plane with respect to the \textit{k}-th grid cell. The expression for the ballistic correction factor is then

\begin{equation} \label{eq:cf_ball}
C\!F_{b}^{ij} = \frac{\bar{A}_{t,av}^{\;ij}}{A_{t,j}}
\end{equation}

Again, as in \cref{subsec:linear_impact}, the correction factors are converted to their linearised counterparts as lengths are used in the standard procedure for the computation of the impact probability (\cref{eq:p_comp_hyp}). The linearised version is again obtained using \cref{eq:d_eq}, giving the following expressions for the correction factors in the hypervelocity and ballistic regimes.

\begin{align}
C\!F_{h}^{ij} &= \frac{\bar{d}_{vz,av}^{ij}}{d_{vz,j}} \label{eq:hyp_cf_lin} \\ 
C\!F_{b}^{ij} &= \frac{\bar{d}_{t,av}^{ij}}{d_{target}^{ij}} \label{eq:ball_cf_lin}
\end{align}

Finally, the hypervelocity and the ballistic correction factors can be applied to the computation of the impact probabilities as follows:

\begin{align}
P_{comp,h}^{\,ij} &= \frac{\bar{d}_{ejecta}^{\,ij}}{l_{vz,j}} \cdot C\!F_{h}^{ij} \label{eq:pc_hyp_cf} \\
P_{comp,b}^{\,ij} &= \frac{C\!F_{b}^{ij} \cdot d_{target}^{\,ij}+d_{p,i}}{l_{vz,j}} \label{eq:pc_ball_cf}
\end{align}

\section{Comparison with DRAMA and ESABASE2/DEBRIS}
\label{sec:comparison}
To verify the validity of the approach described in \cref{sec:vulnerability,subsec:mutual_shielding}, a comparison with two state-of-the-art software packages has been performed. ESABASE2/DEBRIS and \gls{esa} \gls{drama} have been considered. \gls{drama} performs a simplified vulnerability analysis through its dedicated module \gls{midas} \citep{Gelhaus2014,Gelhaus2013,Martin2005,Martin2005a}. The analysis of \gls{midas} is limited to the outer structure of the spacecraft (no internal components can be considered). The user can analyse the debris and meteoroid fluxes and damage for any user-defined target orbit and particle size range. The analysis can be performed for a spherical target, a random tumbling plate, or up to ten oriented surfaces can be defined. The user can select the orientation, area, density, and type of shielding of the panels. There are four hard-coded damage equations and up to 20 can be defined. The debris population uses the fluxes distributions provided by ESA-\gls{master}. The collision flux analysis performed by \gls{midas} provides information about impacting particles and the probability of collision for each of the defined surfaces separately. For this analysis \gls{midas} uses a reference area $S_{\rm ref}$. This area is the cross-section for spherical objects and randomly tumbling plate. For oriented surfaces, the surface areas of each panel are the reference areas. Given the simulation time $\Delta T$, and the impact flux $\varphi$ generated by \gls{master}, \gls{midas} computes the number of impacts $N_{\rm imp}$ as follows

\begin{equation}  \label{eq:midas_nimp}
	N_{\rm imp} = \varphi \cdot S_{\rm ref} \cdot \Delta T
\end{equation}

From which follows the impact probability

\begin{equation}
	P_{\rm imp} = 1 - e^{-N_{\rm imp}}
\end{equation}

The value of $\Delta T$ is computed as the difference between the start and the end epoch defined by the user.
Alongside the impact analysis, \gls{midas} also performs a damage assessment for oriented
surfaces. To do so, it uses the \emph{failure flux} provided by \gls{master}, which is the flux of particles penetrating the surface. Similarly to the collision analysis, the number of penetrations $N_{\rm pen}$ and the penetration probability $P_{\rm pen}$ are computed as follows

\begin{align}
	N_{\rm pen} &= \varphi_{\rm fail} \cdot S_{\rm ref} \cdot \Delta T \\
	P_{\rm pen} &= 1 - e^{-N_{\rm pen}}
\end{align}

The failure flux $\varphi_{\rm fail}$ is generated in a special plug-in routine of \gls{master} where the \glspl{ble} are applied.

ESABASE2/DEBRIS is a more complex software, which allows the user to build an arbitrarily complex structure. The methodology used by ESABASE2 is based on a ray-tracing method, but no debris-cloud propagation is taken into account. However, the vulnerability of internal components can still be analysed by using a particular workaround: the outer structure is \emph{removed} and the user has to provide manually the type of shielding and the stand-off distances for each panel of the internal structure. The presented model, instead, automatically detects the characteristics of the outer structure and assigns the proper stand-off distance and shielding configurations.

\subsection{Test case: Cubic structure in \gls{sso} orbit} \label{subsec:test1}
The first comparison \citep{Trisolini2018_Acta} is aimed at verifying the main building blocks of the model, such as the representation of the environment through \emph{vector flux elements}, the implementation of the ballistic limit equations, and the computation of the impact and penetration probabilities through the approach outlined in \cref{subsec:prob_assessment} and the use of the concept of critical flux. A standard scenario has been selected, where the impact and penetration probabilities are computed for an aluminium Al-6061-T6 cubic-shaped object with 1 m side length and 2 mm thickness. The characteristics of the material are summarised in \cref{tab:Al_properties}. 

\begin{table}[hbt]
\caption{\label{tab:Al_properties} Material properties for aluminium Al-6061-T6.}
\centering
{\renewcommand{\arraystretch}{1.1}
\begin{tabular}{ccccc}
\hline
\bm{$\rho_m$} \bm{$(kg/m^3)$} & \bm{$HB$} & \bm{$C (m/s)$} & \bm{$\sigma_y$} \bm{$(MPa)$} & \\ \hline
2713 & 95 & 5100 & 276 & \\
\hline
\end{tabular}}
\end{table}

The mission considered is a 1-year mission in a \gls{sso} with altitude equal to 802 km, inclination of 98.6 degrees, and eccentricity of 0.001 with starting epoch on the 1$^{\rm st}$ of January 2016. The ballistic limit equation used is the Cour-Palais thin wall \citep{Ryan2010}. The discretisation used for the vector flux elements is summarised in \cref{tab:vfe_bins}. Either the bin location or the number of bins has been specified, alongside the considered boundaries.

\begin{table}[hbt]
\caption{\label{tab:vfe_bins} Binning used for the generation of the \emph{vector flux elements} .}
\centering
{\renewcommand{\arraystretch}{1.2}
\begin{tabular}{c|c|c}
\hline
\bf{Variable} & \bf{Bounds} & \bf{Binning} \\ \hline \hline
$d_p$ & 0.0001 $-$ 0.1 m & 200 \\ \hline
$v_p$ & 0 $-$ 20 km/s & 40 \\ \hline
\multirow{2}{*}{$El$} & \multirow{2}{*}{-90$^\circ$ $-$ 90$^\circ$} & Every 15$^\circ$ in $[$ -90$^\circ$, -30$^\circ$ $)$ $\cup$ $($ 30$^\circ$, 90$^\circ$ $]$ \\ && Every 10$^\circ$ in $[$ -30$^\circ$, 30$^\circ$ $]$ \\ \hline
\multirow{3}{*}{$Az$} & \multirow{3}{*}{-180$^\circ$ $-$ 180$^\circ$} & Every 30$^\circ$ in $[$ -180$^\circ$, -90$^\circ$ $]$ $\cup$ $[$ 90$^\circ$, 180$^\circ$ $]$ \\ && Every 15$^\circ$ in $($ -90$^\circ$, -45$^\circ$ $]$ $\cup$ $[$ 45$^\circ$, 90$^\circ$ $)$ \\ && Every 5$^\circ$ in $($ -45$^\circ$, 45$^\circ$ $)$ \\
\hline
\end{tabular}}
\end{table}

The resulting comparison for the number of impacts and penetration is summarised in \cref{tab:Pimp_comparison} and \cref{tab:Ppen_comparison} respectively.

\begin{table}[hbt]
\caption{\label{tab:Pimp_comparison} Comparison between the numbers of impacts for a cubic structure.}
\centering
{\renewcommand{\arraystretch}{1.1}
\begin{tabular}{lcccc}
\hline
\textbf{Panel} & \textbf{DRAMA} & \textbf{ESABASE} & \textbf{Model} & \\\hline
Lead & 69.473& 80.58& 69.47 & \\
Space & 0.48114 & 2.176 & 1.078 & \\
Trail & 0.032326 & 0.0222 & 0.032 & \\
Earth & 0.54294 & 2.517 & 1.259 & \\
Right & 19.196 & 21.78 & 19.17 & \\
Left & 21.953 & 27.96 & 21.975 & \\
\hline
\hline
Total & 111.678406 & 135.0352 & 112.998 & \\
\hline
\end{tabular}}
\end{table}

\begin{table}[hbt]
\caption{\label{tab:Ppen_comparison} Comparison between the numbers of penetrations for a cubic structure.}
\centering
{\renewcommand{\arraystretch}{1.1}
\begin{tabular}{lcccc}
\hline
\textbf{Panel} & \textbf{DRAMA} & \textbf{ESABASE} & \textbf{Model} & \\\hline
Lead	& 0.2887	& 0.2868	& 0.276 &\\
Space	& 1.71E-05	& 5.65E-05	& 2.12E-05 &\\
Trail	& 7.60E-11	& 5.26E-08	& 1.03E-06 &\\
Earth	& 1.73E-05	& 5.05E-05	& 1.99E-05 &\\
Right	& 0.01027	& 0.0164	& 0.0067 &\\
Left	& 0.01	& 0.0203	& 0.0076 &\\
\hline
\hline
Total	& 0.3090	& 0.3235	& 0.2913 &\\
\hline
\end{tabular}}
\end{table}

A good agreement is observed between the developed model and \gls{drama}, with results that differ from about 1 impact and 0.1 penetrations per year, which are comparable to other analyses \citep{Miller2017}. The results of the ESABASE2 simulations instead show a higher number of both predicted impacts and penetrations with respect to the other two software packages. This difference may be explained with ESABASE2 using a different \gls{master}-2009 population. Nonetheless, the higher number of penetrations in ESABASE2 can then be directly connected to the higher number of impacts. The difference with \gls{drama} is not substantial but still deserves an analysis. In fact, this is focused only on two of the faces (Earth and Space), while the others have almost identical results. This difference can be traced back to the discretisation employed in the generation of the \emph{vector flux elements} (\cref{tab:vfe_bins}) combined with the strongly directional nature of the fluxes in \gls{sso} orbits. The impacts on these two faces are mainly a function of the impact elevation angle. As this has a stronger directionality than the azimuth, it is more susceptible to the discretisation used. Given the adopted discretisation in the impact elevation, it may not be able to predict the fluxes on the Earth and Space faces as well as for the other ones.

\subsection{Test case: Single child component in \gls{sso} orbit} \label{subsec:test2}
\begin{figure}[htb!]
 \centering
 \includegraphics[width=0.65\textwidth, keepaspectratio]{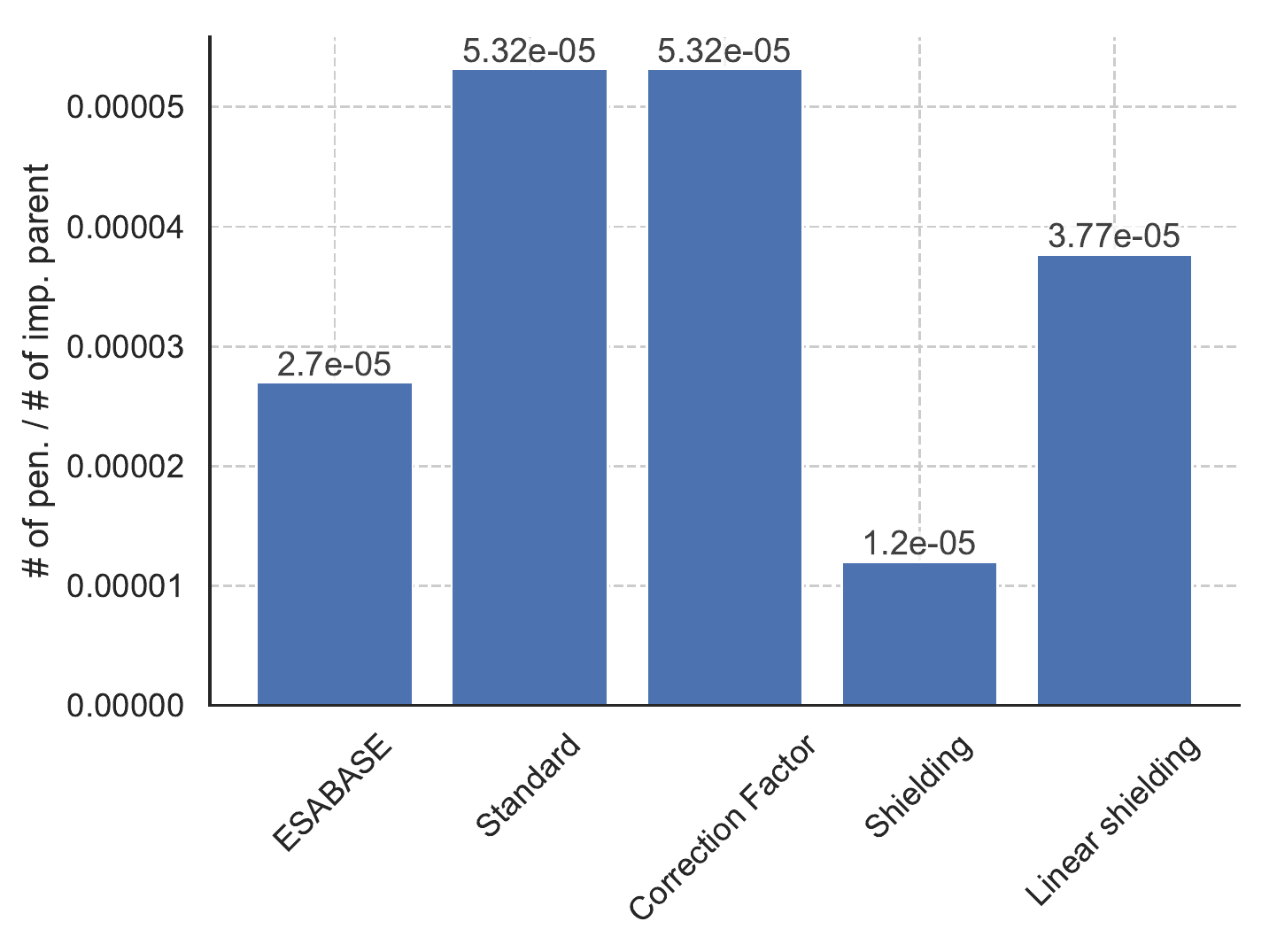}
 	\caption{Comparison for one internal component.}
 	\label{fig:comparison_one}
 \end{figure}

Once the general procedure has been verified with a standard test case against \gls{drama} and ESABASE2/DEBRIS, the model has been tested also for the computation of the vulnerability of internal components. For this test case, only a comparison with ESABASE2/DEBRIS could be performed as \gls{drama} does not support such an analysis. First, the model is tested for a configuration with a single internal component. A cubic-shaped parent object made of aluminium Al-6061-T6 with a 1 m side length and a 1 mm wall thickness has been selected. The child component contained in the parent is box-shaped, made of the same material, with a side length of 40 cm, and a wall thickness of 1 mm. The mission scenario is the same as the previous test case (\cref{subsec:test1}). The ballistic limit equation used is the ESABASE Double Wall \citep{Gade2013}. \cref{fig:comparison_one} shows the results of the comparison between ESABASE2/DEBRIS and different options of the developed model. The different options presented are the standard vulnerable zone approach (\cref{sec:vulnerability}) identified with the label \emph{Standard}, the correction factor methodology (\cref{subsec:correction_factor}), the impact ejecta methodology considering the areas (\cref{subsec:hypervelocity,subsec:ballistic}) identified by the label \emph{Shielding}, and the linearised version of the impact ejecta method (\cref{subsec:linear_impact}) identified by the label \emph{Linear shielding}. Given the difference between the impacts predicted by this model and ESABASE2/DEBRIS (\cref{tab:Pimp_comparison}), it was decided to compare the ratio between the number of penetrations and the number of impacts on the outer structure, instead of directly comparing the number of penetrations.

\cref{fig:comparison_one} show that all the models predict a penetration ratio that is of the same order of magnitude. The model predicting the largest amount of penetrations is the standard vulnerable zone methodology. This is expected as this methodology is an intrinsically conservative procedure. Also expected is the behaviour of the correction factor methodology, which replicates the results of the standard formulation as no shielding components are present in the test case, resulting in a value of 1 for the correction factors of \cref{eq:hyp_cf_lin,eq:ball_cf_lin}. The two versions of the ejecta model instead exhibit quite a different behaviour, with the linearised version showing the best agreement with ESABASE2/DEBRIS, while the version with the areas clearly gives the lowest result. This difference can be expected: both models rely on the same computational procedure to evaluate the interactions between the debris ejecta and the shielding components; however, the linearisation process increases the computed value of $P_{cloud}$ as the ratio between the areas will always be lower than the ratio between the associated equivalent lengths.

\subsection{Test case: Two children components in \gls{sso} orbit} \label{subsec:test3}
\begin{figure}[htb!]
 \centering
 \includegraphics[width=0.6\textwidth, keepaspectratio]{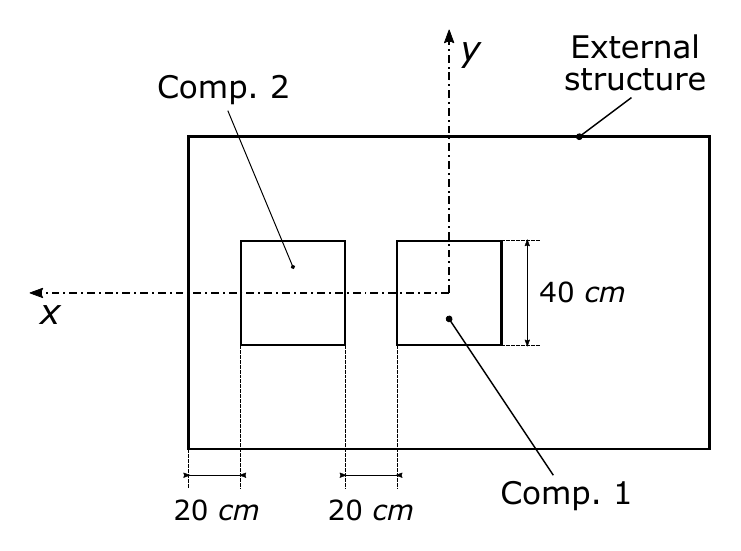}
 	\caption{Schematics of the satellite geometry used for the mutual shielding comparison with ESABASE2/DEBRIS.}
 	\label{fig:comparison_two_draw}
\end{figure}

For the final comparison, a parent structure with two internal components is considered. The parent structure is a $2m \times 1m \times 1m$ aluminium Al-6061-T6 box with a 1 mm wall thickness. The two internal components are two identical box-shaped objects with a 40 cm side length and a 1 mm wall thickness. The first box (\textit{Component 1}) is in the centre of the parent structure, while the second box (\textit{Component 2}), is positioned in front of the first one along the RAM direction at a distance of 0.2 m from the outer face (\cref{fig:comparison_two_draw}). Again, the mission scenario is equivalent to the one of \cref{subsec:test1}.

\begin{figure}[htb!]
 \centering
 \includegraphics[width=0.65\textwidth, keepaspectratio]{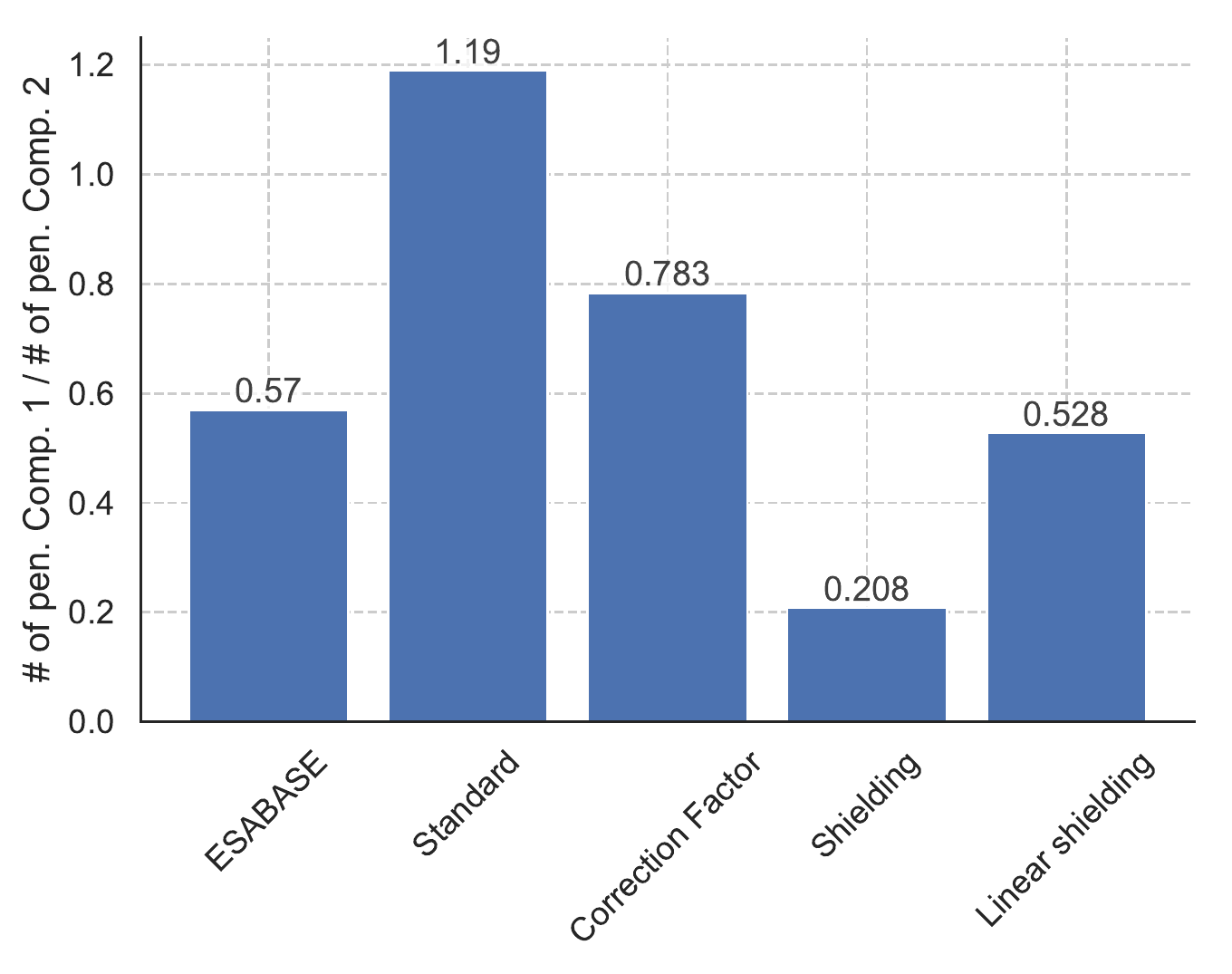}
 	\caption{Comparison for two internal components.}
 	\label{fig:comparison_two}
\end{figure}

This comparison is used to test the capability of the code to deal with shielding components (\cref{subsec:mutual_shielding}). For the comparison, the ratio between the penetrations on the lead face of \textit{Component 1} and \textit{Component 2} is presented in \cref{fig:comparison_two}.

\cref{fig:comparison_two} shows that, as expected, the standard formulation is not able to correctly model the vulnerability of internal components when shielding needs to be considered. In fact, \textit{Component 1} has even more penetrations than \textit{Component 2}. In this methodology, the two components are treated separately and \textit{Component 1} receives more penetration because the overall extent of the vulnerable zone is larger than the one of \textit{Component 2} as it is positioned further away from the external face of the structure. The remaining three methodologies, instead, are all capable of comping with mutual shielding. The results of \cref{fig:comparison_two} shows that the correction factor model is the most conservative as it predicts a higher fraction of penetration between the two components. Similarly to \cref{fig:comparison_one}, the ejecta model using the areas (\emph{Shielding}) is the least conservative and predicts a considerably lower amounts of penetrations for \textit{Component 1}. Finally, the linearised version of the ejecta model more closely matches the results of ESABAS2/DEBRIS.
An interesting aspect of both the comparisons presented in \cref{fig:comparison_one,fig:comparison_two} is the significant difference between the procedure using area intersection for the assessment of the mutual shielding and the correspondent linearised version. As previously mentioned, this behaviour is expected. What is less expected is the better agreement of the linearised version with state-of-the-art software packages, rather than a \emph{more natural} implementation based on the intersection of sections. This suggests that an extent-based computational methodology should be favoured over an area-based one for the implementation of a vulnerable zone-based computation of the vulnerability. This is in agreement with the idea of this work that is to extend the original vulnerable zone methodology, whose probability assessment ultimately relies on an extent-based procedure.

\subsection{Execution time}
This section contains a summary of the execution time of the presented methodology for the test cases presented in \cref{subsec:test1,subsec:test2,subsec:test3} (\cref{tab:exe_time}). The tests have been run on a system equipped with an i7-8700 CPU @ 3.2 GHz with 6 cores and 16 GB of RAM. Our model uses pre-computed flux distributions from \gls{master}, and its execution time is not included in the runtime.

\begin{table}[hbt]
\caption{\label{tab:exe_time} Execution time for the presented test cases (average over 10 simulations).}
\centering
{\renewcommand{\arraystretch}{1.1}
\begin{tabular}{lcc}
\hline
\textbf{Test case} & \textbf{Runtime} \\ \hline
Test 1 & 0.53 s \\
Test 2 - Correction factor & 0.59 s \\
Test 2 - Linear shielding \footnotemark & 1.16 s \\
Test 3 - Correction factor & 0.69 s \\
Test 3 - Linear shielding & 5.56 s \\
\hline
\end{tabular}}
\end{table}
\footnotetext{The execution time for the \emph{Shielding} and \emph{Linear shielding} cases are equivalent}

\section{Conclusions and Discussion}
\label{sec:conclusions}
The paper has presented the development of a novel methodology to predict the vulnerability of spacecraft configurations to the impact with debris particles, with particular attention to devising a methodology for taking into account mutual shielding effects between internal components. The method has its foundations in the concept of the vulnerable zone and builds on it to obtain a statistically robust procedure for survivability assessments. A complete description of the main building blocks and methodologies implemented in the model has been given. A novel procedure for the computation of the mutual shielding contribution to the damage to internal components has been presented, based on the geometrical interaction between the secondary debris ejecta, modelled as conic shapes and the shielding and target components. The overall methodology has been tested and compared with state of the art software packages, showing very good agreement with traditional impact assessment methodologies. Such a comparison served also as a verification test for the fully probabilistic approach adopted, where the penetration probability is computed as the product of the probabilities of three separate events. Also, the mutual shielding capabilities have been compared with ESABASE2/DEBRIS, showing comparable results and demonstrating that it is possible to predict the mutual shielding between components using the interaction of geometrical shapes and avoiding ray tracing methodologies.

Alongside the obtained results, it should be mentioned that the presented methodologies is in its initial level of development and has been verified against limited test cases and scenarios. Consequently, for future development, a more complete set of tests should be executed, including more complex internal configurations, with an increasing number of components and different shapes. This is especially useful for verifying the adaptability and scalability of the mutual shielding procedure to more complicated geometries. Moreover, different shielding methodologies should also be tested, specifically including honeycomb sandwich panels as they are among the most used spacecraft structures.

\section*{Acknowledgements}
This work was funded by EPSRC DTP/CDT through the grant number EP/K503150/1.

\bibliographystyle{elsarticle-harv}
\bibliography{bibliography}

\appendix
\section{Material database}
\label{sec:srl_coeff}
In \cref{tab:srl_coeff} are summarised the values of the coefficients for the \gls{srl} \gls{ble} for the two different cases of Aluminium outer bumper plate and \gls{cfrp} outer bumper plate \citep{Christiansen2009}.
\begin{table}[!hbt]
\caption{Summary of the coefficients for the \gls{srl} \gls{ble}}
\label{tab:srl_coeff}
\centering
{\renewcommand{\arraystretch}{1.1}
\begin{tabular}{lccc}
\hline
\textbf{Symbol} & \textbf{Aluminium outer bumper} & \textbf{\gls{cfrp} outer bumper} & \\\hline
$V_{LV}$ & 3 km/s & 4.2 km/s & \\
$V_{HV}$ & 7 km/s & 8.4 km/s & \\
$K_{3S}$ & 1.4 & 1.1 & \\
$K_{3D}$ & 0.4 & 0.4 & \\
$K_{tw}$ & 1.5 & 1 & \\
$K_{S2}$ & 0.1 & 1 & \\
$\beta$ & 2/3 & 1/3 & \\
$\delta$ & 
	\begin{tabular}{@{}c@{}} 4/3 \text{if} $45^{\circ} \geq \theta \geq 65^{\circ}$ \\ 4/3 \text{if} $45^{\circ} \leq \theta \leq 65^{\circ}$ \end{tabular}
 	& 4/3 & \\
$\epsilon$ & 
	\begin{tabular}{@{}c@{}} 8/3 \text{if} $45^{\circ} \geq \theta \geq 65^{\circ}$ \\ 10/4 \text{if} $45^{\circ} \leq \theta \leq 65^{\circ}$ \end{tabular}
	& 0& \\
$\gamma$ & 1/3 & 2/3 & \\
\hline
\end{tabular}}
\end{table}

\end{document}